# DiSE: A diffusion probabilistic model for automatic structure elucidation of organic compounds

Haochen Chen[1], Qi Huang[1], Anan Wu[1]*, Wenhao Zhang[2], Jianliang Ye[1], Jianming Wu[2], Kai Tan[1], Xin Lu[1]*, Xin Xu[3]*

[1]State Key Laboratory of Physical Chemistry of Solid Surfaces and Fujian Provincial Key Laboratory of Theoretical and Computational Chemistry, College of Chemistry and Chemical Engineering, Xiamen University, Xiamen, China.
[2]Shanghai Key Laboratory of Electrochemical and Thermochemical Conversion for Resources Recycling, Department of Chemistry, Fudan University, Shanghai, China.

[3]Collaborative Innovation Center of Chemistry for Energy Materials, Shanghai Key Laboratory of Molecular Catalysis and Innovative Materials, MOE Key Laboratory of Computational Physical Sciences, Department of Chemistry, Fudan University, Shanghai, China.

*Corresponding author. Email: ananwu@xmu.edu.cn (A.W.); xinlu@xmu.edu.cn (X.L.); xxchem@fudan.edu.cn (X.X.)

**Abstract:** Automatic structure elucidation is essential for self-driving laboratories as it enables the system to achieve truly autonomous. This capability closes the experimental feedback loop, ensuring that machine learning models receive reliable structure information for real-time decision-making and optimization. Herein, we present DiSE, an end-to-end diffusion-based generative model that integrates multiple spectroscopic modalities, including MS, $^{13}$C and $^{1}$H chemical shifts, HSQC, and COSY, to achieve automated yet accurate structure elucidation of organic compounds. By learning inherent correlations among spectra through data-driven approaches, DiSE achieves superior accuracy, strong generalization across chemically diverse datasets, and robustness to experimental data despite being trained on calculated spectra. DiSE thus represents a significant advance toward fully automated structure elucidation, with broad potential in natural product research, drug discovery, and self-driving laboratories.

## 1. Introduction

Modern self-driving laboratories (SDL), also known as autonomous laboratories, are revolutionizing chemical research by integrating robotics, artificial intelligence (AI), and machine learning (ML) to accelerate research in chemistry and material discovery (*1–11*). These systems aim to make the entire process of chemical synthesis (e.g., organic synthesis) – from experimental design, sample preparation, synthesis, in-line characterization and data-driven decision-making – highly automated or even completely autonomous (*9*). Achieving such a close-loop framework requires synergistic advances across multiple disciplines, including automation technologies (*5*), AI-driven experimental design (*6*) and the integration of diverse analytical techniques (*7, 8*). Among these, automatic structure elucidation represents one of the most critical and challenging components, as it plays a crucial role in validating products, quantifying outcomes, catching errors, and guiding subsequent experiments (*9–11*).

Computer-aided structural elucidation (CASE) was developed for this purpose, which automates the determination of molecular structures from spectroscopic data (*12*), including MS, one- and two-dimensional NMR (1D/2D NMR), and/or IR spectroscopy. These methods typically compare experimental spectra against reference databases to generate candidate structures, which are then ranked using forward-prediction models (*12–15*). Recent developments have incorporated density functional theory (DFT) calculations and advanced statistical models to improve performance (*16–24*). While effective for well-characterized molecular frameworks, CASE approaches often struggle with novel or hydrogen-deficient compounds due to sparse reference data and ambiguous spectral features, which limit both accuracy and scalability (*12*, *15*). Furthermore, traditional CASE methodologies typically require significant human intervention (*25*) and depend heavily on the precision of forward-prediction models/methods (*12–15*, *26*, *27*). Despite the rich structural information encoded in spectroscopic data—particularly NMR parameters—current database- and DFT-driven CASE strategies have yet to fully leverage the inherent spectrum–structure relationship.

The core challenge in automatic structure elucidation is to accurately determine the correct atomic and functional group connectivity within the vast chemical space (*28–30*) that matches the given spectral characterization data. This task becomes increasingly difficult as molecule size increases, since the number of possible structures grows combinatorially (*31, 32*), far exceeding the capacity of exhaustive or brute-force enumeration. Comparable difficulties arise in areas such as protein structure prediction (*33–35*), protein conformations (*36*), and inorganic material design (*37*) —domains where deep generative models have recently shown considerable promises. These models offer a more efficient and creative strategies for chemical space exploration than traditional enumeration methods (*16–24*, *38*, *39*). By directly learning the correlation between spectra and molecular structures, generative models can capture intra- and inter-spectral dependencies from multimodal spectra and reframe structure elucidation as a probabilistic mapping from a continuous spectral domain to a discrete molecular graph, thereby providing a powerful, scalable, and end-to-end approach for the automatic structure determination of organic molecules.

In recent years, several research groups (*25*, *28*, *29*, *40–46*) have explored deep generative models for automatic structure elucidation of organic compounds [see supplementary materials (SM) Table S1 for details]. These methods typically treat molecules as token sequences and employ Transformer trained on the Simplified Molecular Input System (SMILES). Despite promising progress, SMILES-based approaches face intrinsic limitations: their text-like format cannot effectively encode atomic correlation and integrate 2D spectral information. Consequently, such models fail to incorporate key 2D NMR modalities — particularly homonuclear correlation spectroscopy ($^3J_{HH}$ COSY) and heteronuclear single quantum coherence (HSQC) — which provide essential connectivity information for structure elucidation (*12*, *25*). Neglecting these spectra often results in incorrect structure. Moreover, models trained on low-quality synthetic data in these methods may suffer from limited generalization capabilities (*47*).

2D NMR spectra, such as HSQC and COSY, contain matrix-like correlations that are difficult to tokenize or normalize for natural language processing (NLP) style models. Incorporating such data necessitates a 2D molecular representation, such as 2D molecular graph. Notably, HSQC and COSY are not only highly informative but also experimentally efficient, typically requiring shorter acquisition times than $^{13}$C NMR. Thus, integrating $^1$H, $^{13}$C NMR, COSY and HSQC data can significantly enhance predictive accuracy while maintaining experimental feasibility—an essential feature for real-time operation in SDLs.

Herein, we introduced DiSE, an end-to-end deep generative framework that leverages graph-based molecular representations and discrete diffusion (*48*) modeling to generate molecular structures directly from spectroscopic data. DiSE embodies the probabilistic mapping paradigm outlined above: instead of generating and evaluating candidate structures separately, it models the relationship between spectra and molecular structures as a continuous-to-discrete probabilistic transformation. By learning this mapping through machine learning and data-driven approach, DiSE can achieve automatic structure elucidation without relying on exhaustive enumeration or computationally expensive quantum chemical calculations. We demonstrated, using a high-precision DFT-calculated NMR chemical shifts dataset comprising 1.7 million molecules, that DiSE could accurately and efficiently determine the structure of organic compounds based on molecular formula, $^{13}$C and $^1$H NMR data, COSY, and HSQC spectra. Notably, the generative diffusion process employed by DiSE exhibits characteristics similar to the heuristic reasoning used by expert chemists, achieving a balance between prediction accuracy and interpretability. With its advantages of speed, interpretability, and adaptability, DiSE can serve as an important component of modern

SDLs, enabling autonomous structure elucidation and real-time structure verification in closed-loop discovery workflows.

## 2. Results and Discussion

### 2.1 Designing and training the diffusion model

In chemistry, organic molecules can be naturally depicted as 2D graphs in which atoms correspond to nodes and bonds to edges. With a molecular formula obtained from MS and $CH_x$ (x=0–3) fragments identified by HSQC, structure elucidation becomes a task of assigning the correct edges between nodes, which are chemical-informed with $^{13}C/^{1}H$ chemical shifts and $^{3}J_{HH}$ COSY, in this graph. To solve this inverse problem, we introduce **DiSE**, a graph-based diffusion model designed specifically for organic molecules. DiSE follows a two-stage workflow (Fig. 1a)—edge-noise injection and subsequent denoising—and employs a tailored graph representation that fully leverages all available spectra, particularly $^{3}J_{HH}$ COSY correlations. Each molecule is encoded as an undirected graph *G*,

$$G = (X, E, E_{COSY}, Y)$$

where *X* contains node features, *E* edge features, $E_{COSY}$ COSY-specific edges, and Y global attributes such as the normalized diffusion timestep (Fig. 1b; SM section A.1.).

#### 2.1.1 Noise-injection stage

A forward Markov chain progressively corrupts the edge tensor $E_0$, producing the sequence ($E_1$, $E_2$, ..., $E_{Tmax}$). Thus, $E_0$ is associated with the true molecular graph $G_0$, whereas $E_{Tmax}$ corresponds to fully corrupted graph $G_{Tmax}$ (Fig. 1a and 1b, SM section A.2.1.). This formulation (i) allows any intermediate state $E_t$ to be sampled directly from $E_0$ without storing the entire chain, and (ii) ensures local reversibility in the reverse process, so the graph neural network (GNN) need learn only a single-step denoising map to achieve global optimization. Empirically, $T_{max}$=500 with a cosine-annealed $\beta_t$ schedule balances noise robustness and training stability.

#### 2.1.2 Denoising stage

DiSE employs a custom graph transformer (GT) architecture (Fig. 1b), which transmits long-range information more efficiently than conventional message-passing neural networks (MPNNs) (*49*). The network comprises 20 residual GT blocks with 32 attention heads each. Node features *X* are embedded in a 1,024-dimensional space, whereas the concatenated tensor $E_t \oplus E_{COSY}$ is embedded in 512 dimensions. Within each block, multi-head attention updates node embeddings using the edge information, followed by layer normalization, a residual connection, and a feed-forward network (Fig. S1 and Table S2). Edge features $E_t$ are updated in parallel, with weights conditioned on the current timestep $t$. Because $X_{Spectra}$ and $E_{COSY}$ are time-invariant whereas $E_t$ evolves, this "static–dynamic" decoupling markedly enhances the model's ability to learn spectrum–structure correlation.

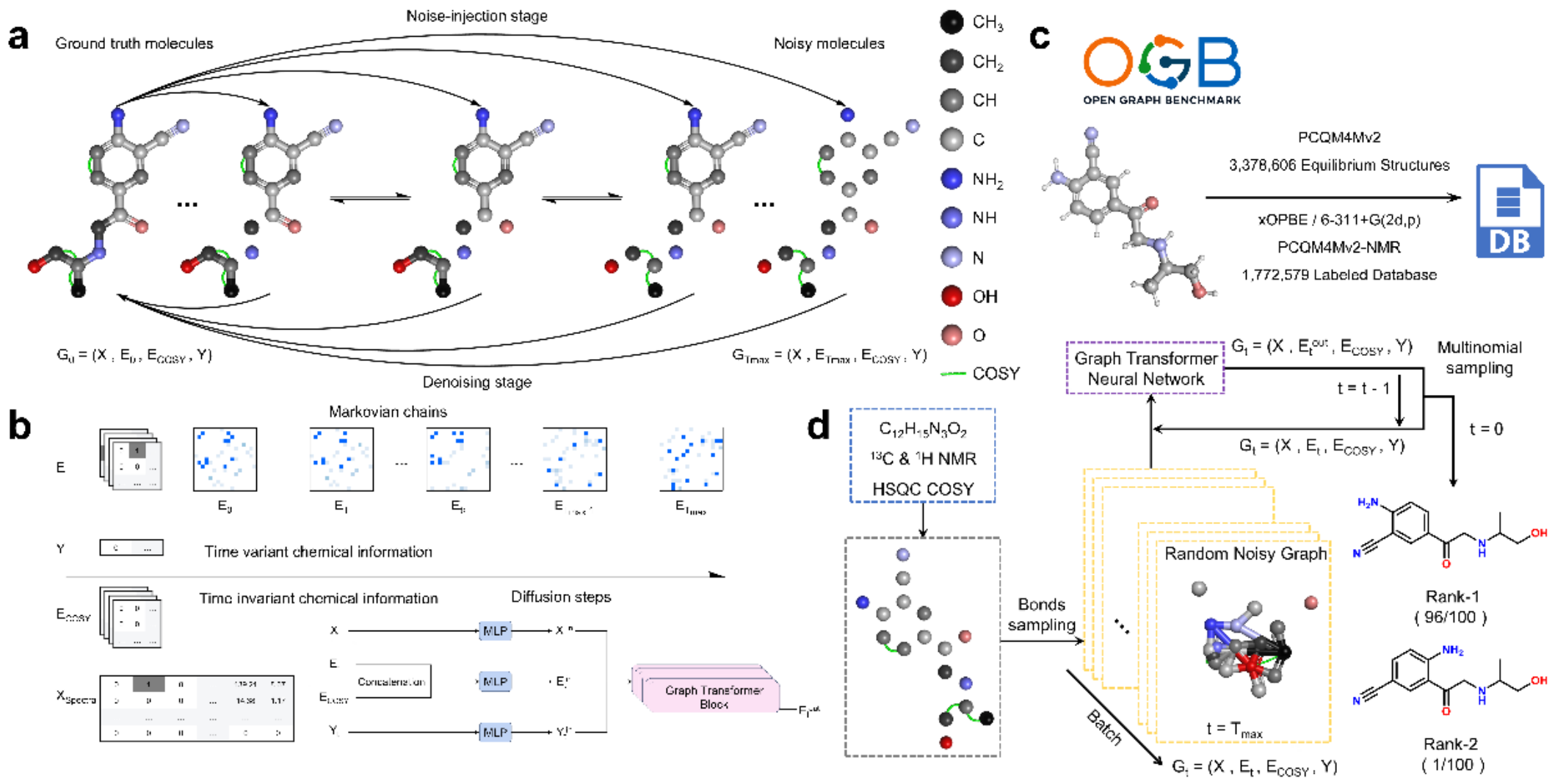

**Fig. 1. Schematic diagram of DiSE a.** Two-stage workflow employed in DiSE: a forward noise injection stage, which perturbs the truth molecular graph ($G_0$) into the noisy state ($G_t$), and the reverse denoising stage, which reconstructs the original molecules from the fully corrupted graph ($G_{Tmax}$) using a neural network. The colored balls represent super-atom types ($XH_n$, X: C, O, N and n: 0-3). **b.** Denoising stage. Time-variant chemical information ($Y_t$) with time-invariant features ($X_{Spectra}$) and ($E_{COSY}$) are processed through a graph transformer to predict the truth molecular graph from the previous stage (see supplementary materials for details). **c.** Construction of the high-quality NMR chemical shifts dataset based on xOPBE (PCQM4Mv2-NMR). **d.** The overall inference workflow. During inference, DiSE generates random molecular graphs as input by multinomial sampling and then iterative denoising to generate a ranked list of candidate molecular structures. The rankings were based on their frequency of occurrence.

### 2.1.3 Model Training and Inference

We evaluated the reliability and generalizability of the DiSE model using two molecular datasets containing only C, H, O and N atoms. First, we employed the cleaned QM9-NMR benchmark dataset (*50*), comprising 116,977 small molecules with mPW1PW91-calculated NMR properties, to assess model performance in small-molecule structure elucidation. To ensure the applicability and higher generalization ability of the model in a broader and more realistic chemical spaces, we constructed an extended dataset, PCQM4Mv2-NMR (Fig. 1c), consisting of 1,772,579 molecules. This dataset was derived from the conformational equilibrium molecular library provided by the Open Graph Benchmark (OGB) (*51*), where the $^{13}C$ and $^1H$ chemical shifts were calculated using the more accurate functional xOPBE (*52*). To ensure data quality, we filtered out molecules from original PCQM4Mv2 that (i) exhibited bond-order inconsistencies identified by RDKit (*53*), or (ii) contained NMR chemical shifts outside typical detection ranges (>15 ppm for $^1H$ or >250 ppm for $^{13}C$). Both datasets were randomly partitioned into training, validation, and test sets using an 8:1:1 split, ensuring no data leakage.

DiSE was trained to process multiple spectral inputs – MS, $^1$H, $^{13}$C, HSQC and $^3J_{HH}$ COSY – to predict molecular connectivity. During the training, the model minimized the negative log-likelihood (NLL) of edge-type predictions using the AdamW optimizer with a learning rate of 2e$^{-3}$. During the inference (Fig. 1d), molecular structures were generated via a reverse Markov chain: starting from a fully corrupted edge tensor $E_{Tmax}$, sampled from a prior distribution of bond types estimated from the training data (Table S2), the model iteratively denoised the graph through $T_{max}$ steps to recover $E_0$. The output of DiSE is a set of constitutional isomers. To promote candidate diversity, 100 independent inference runs were performed per target unless otherwise specified, and the Top-K (e.g., K=1, 3, 5) structures were selected based on their frequency of occurrence across runs.

## 2.2 Model Analysis

To evaluate the performance of DiSE, we adopted a Top-K accuracy metric, which is widely used in structure elucidation (*28*, *29*, *54*) and retrosynthesis (*55–57*). Unlike similarity-based metrics employed in other studies (*25*, *38–40*), we define success strictly as the true structure appearing within the Top-K predicted structures.

On the QM9-NMR test set, DiSE demonstrated strong predictive performance, substantially surpassing previously reported models (Fig. 2a). A single tuned DiSE model achieved Top-1 and Top-3 accuracies of 92.76% and 96.78%, respectively. Incorporating an ensemble strategy further improved performance to 93.59% and 98.22% (Table S3). For comparison, the best previously reported model (MST ref. *43*) reached only 73.38% Top-1 accuracy. DiSE maintained high accuracy on the more chemically diverse PCQM4Mv2-NMR test set, with Top-1 and Top-3 accuracies of 92.11% and 96.41%, which improved to 92.55% and 96.95% with ensemble. In comparison, the NLP-based MMST (*25*) method—despite using similar spectral inputs—achieved only 51.0% Top-1 accuracy. While a similarity-based metric superficially boosted the reported accuracy of MMST to 82.0% (Fig. 2a), such metrics are inappropriate for structure elucidation, where identifying the exact molecular structure—not a similar one—is essential. These results underscore DiSE's superior accuracy, generalization, and transferability across chemically diverse datasets.

To assess the relative contribution of each spectral modality, we conducted ablation studies on the QM9-NMR dataset using various combinations of spectroscopic inputs (SM section A.1.). Importantly, the molecular formula derived from MS is indispensable for structure elucidation, as its absence substantially increases the risk of mis-assignment (*12*). As shown in Fig. 2d, HSQC emerged as the most critical spectral input for model performance, with its omission causing a sharp decline in Top-1 accuracy from 80.68% to 34.93% (Table S4), highlighting its central role in identifying $CH_x$ fragments and drastically narrowing the sampling chemical space. This observation is consistent with both molecular fingerprint theory and previous findings (*12*, *25*, *58*). Combining HSQC with 1D spectra ($^{13}C$ and $^{1}H$ chemical shifts) yielded Top-3 and Top-5 accuracies of 90.34% and 91.70%, respectively. Adding COSY further improved performance to 90.66% (Top-1), 96.26% (Top-3), and 96.76% (Top-5). Additionally, combining HSQC and $^{1}H$ chemical shifts enabled the model to infer exchangeable protons (e.g., hydroxyl and $NH_x$ groups), resulting in a Top-1 accuracy of 92.22%. Interestingly, while only 1D spectra alone yielded poor Top-1 accuracies (20.19% for $^{13}C$ and 13.93% for $^{1}H$), they remained useful, with Top-all accuracies reaching 61.03% for $^{13}C$ and 63.55% for $^{1}H$ (Table S4). Collectively, these results highlight the essential role of 2D NMR data (HSQC and COSY) in accurate structure elucidation and provide practical guidance for optimizing experimental data acquisition strategies.

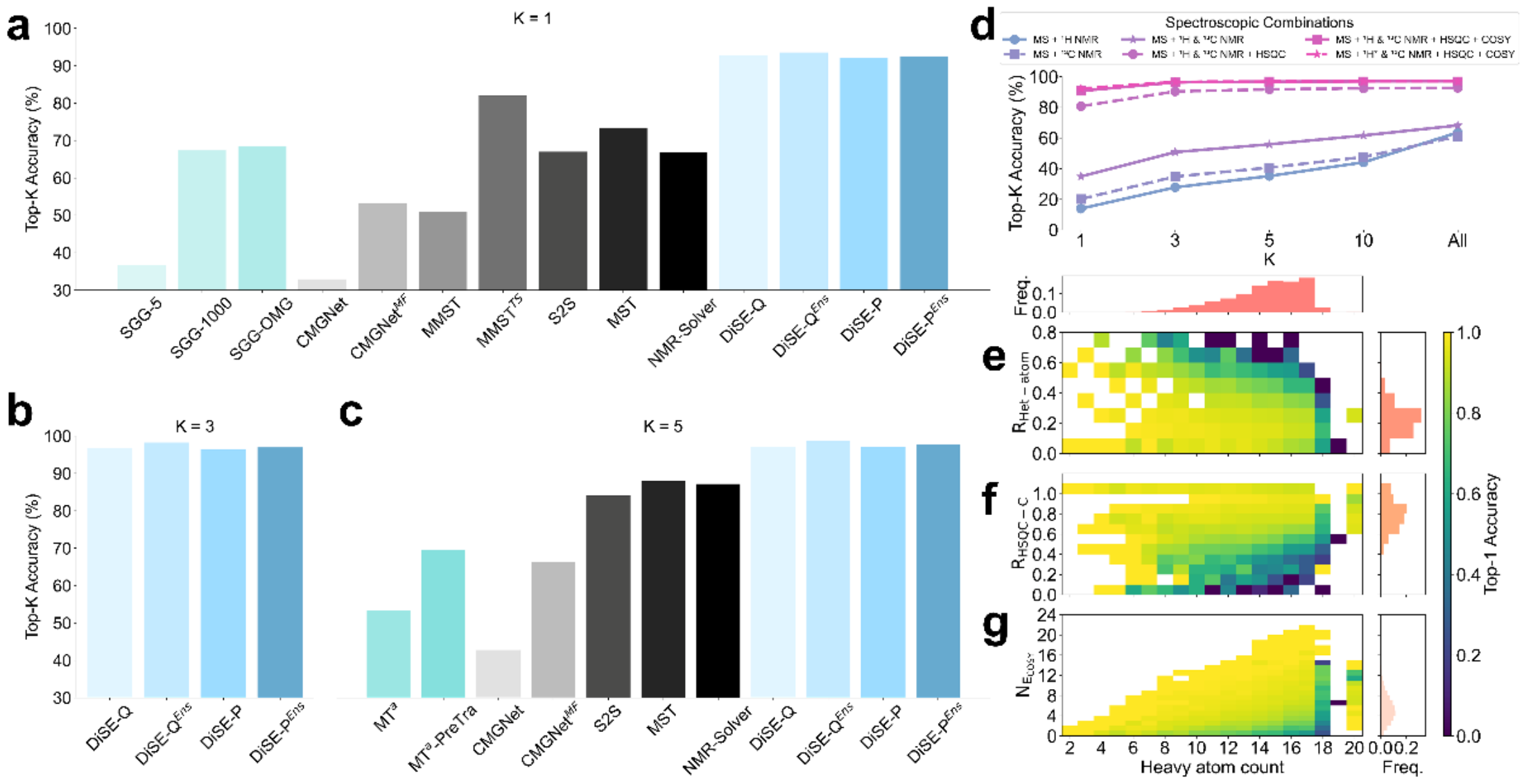

**Fig. 2. Performance and model analysis of DiSE. a.** Comparison of Top-1 accuracy of various models (see Table S1 for method abbreviations, SGG-5, SGG-100 and SGG-OMG from ref. *28*, CMGNet from ref. *40*, MMST from ref. *25*, S2S from ref. *41*, MST from ref. *43*, NMR-Solver from ref. *38*. DiSE-Q: based on the QM9-NMR dataset; DiSE-P: based on the PCQM4Mv2-NMR dataset; superscript *Ens* refers to ensemble models, superscript *MF* refers to models with molecular formula filter, superscript *TS* refers to models based on the Tanimoto Similarity metric). **b.** and **c.** Comparison of Top-3 (**b**) and Top-5 (**c**) accuracy of various models. **d.** Ablation study results for DiSE model. (superscript *: exchangeable protons information used). **e.** to **g.** 2D histograms showing Top-1 accuracy (color scale) as a function of molecular size represented by the number of heavy atoms (molecular weight data also available in Fig. S4) and specific features: **e.** Ratio of heteroatoms, **f.** Ratio of HSQC-carbon atoms, and **g.** the number of $E_{COSY}$. Top and side histograms present the data distribution for each axis.

To further examine DiSE's applicability, we explored the boundaries of structure plausibility by analyzing the impact the molecular size represented by the number of heavy atoms, HSQC/COSY coverage, and heteroatom content on DiSE accuracy. As illustrated in Figs. 2e-2g, DiSE accuracy generally deceased with increasing molecular size. Accuracy also strongly correlated with HSQC (Fig. 2f) and COSY (Fig. 2g) coverage: higher HSQC and COSY coverage can effectively reduce the initial structure ambiguity and hence improve DiSE performance. Notably, heteroatoms (Fig. 2e) pose unique challenges within the DiSE framework because they act as "silence nodes" (*12*). That is, beyond atomic identify, no direct spectral ($X_{Spectra}$) or edge-related information is assigned to heteroatoms; their bonding environment must be inferred indirectly from adjacent carbon atoms. Consequently, molecules with high heteroatom content are particularly difficult to resolve (Fig. 2e). Incorporating heteroatom-specific information, such as IR and Raman spectra, into future model will likely improve performance.

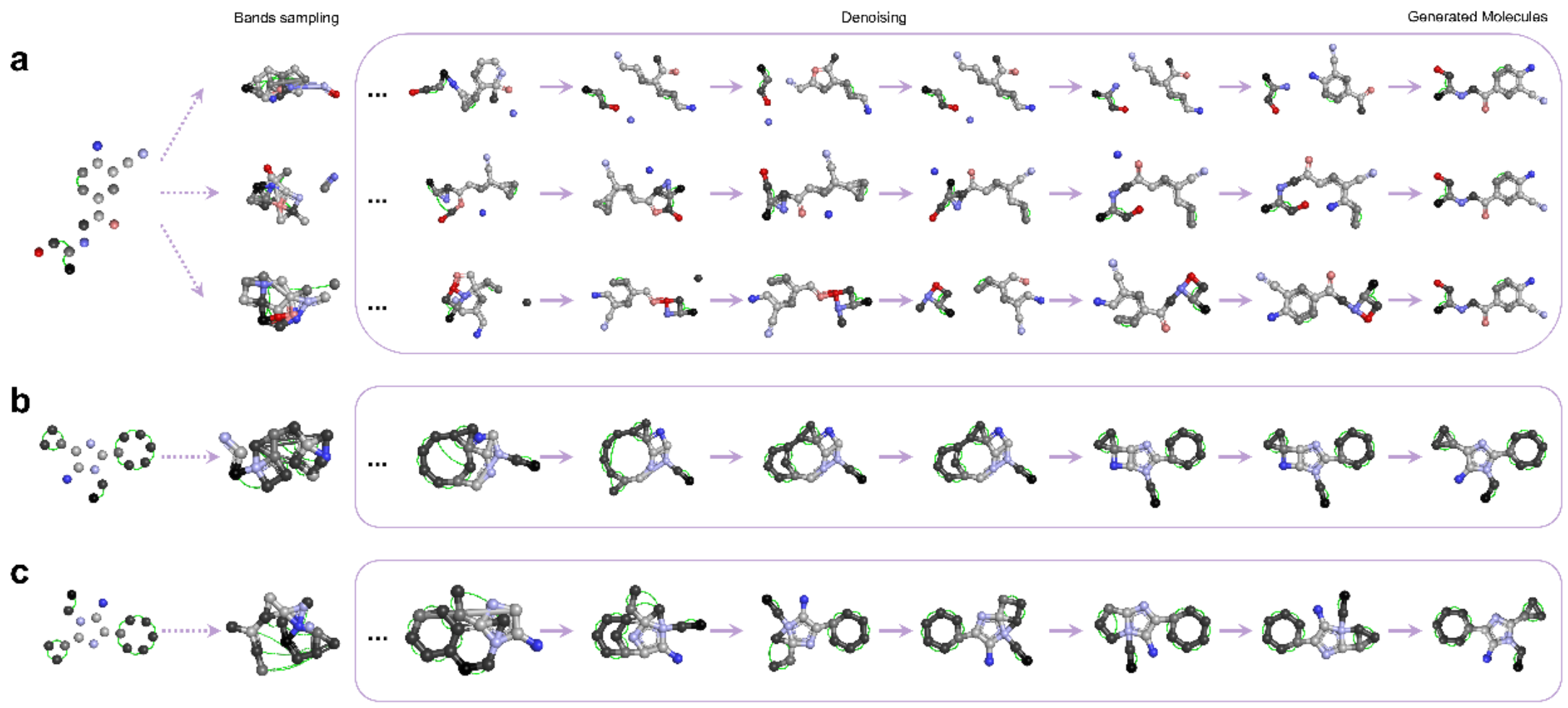

**Fig. 3. Visualization of selected denoising paths.** The inference begins with bonds sampling from a priori distributions to create initial noised molecular graphs (left). And then subjected to a denoising process (center), which is an iterative refinement trajectory that progressively corrects the molecular structure. Eventually, valid candidate molecular structures are deduced (right). **a.** Visualization of denoising trajectories for 2-amino-5-[2-(1-hydroxypropan-2-ylamino)acetyl]benzonitrile, initiated with different random seeds. **b.** and **c.** Visualization of denoising trajectories of positional isomers 2-cyclohexyl-5-cyclopropyl-3-ethylimidazol-4-amine (**b**) and 5-cyclohexyl-2-cyclopropyl-3-ethylimidazol-4-amine (**c**).

We also assessed the interpretability by visualizing denoising trajectories. Fig. 3a illustrates three denoising paths for 2-amino-5-[2-(1-hydroxypropan-2-ylamino)acetyl]benzonitrile, each initiated with different random seeds at $T_{max}$. Unlike NLP-based methods, diffusion models generated structures through iterative denoising of sampled noise, thereby mimicking physical processes (*59–61*). Consistent with expert reasoning, DiSE first generated nodes guided by MS, 1D chemical shifts, and HSQC, then used COSY to establish bonds between adjacent carbon atoms ($E_{COSY}$, green lines in Fig. 3) and gradually clarified $E_{COSY}$-related bond orders, which might be single, double or aromatic bonds, during the denoising process. This further emphasizes the crucial role of 2D spectra in resolving complex structures (*12*, *25*). Bonds involving heteroatoms were typically the most difficult and often resolved last, requiring multiple

iterations – a challenge consistent with the inherent complexity of silent nodes. DiSE also demonstrated the ability to distinguish structurally similar molecules even when their Tanimoto similarity (*62*) was 1, a task that is particularly challenging for similarity-based models. For example, Fig. 3b and 3c show that DiSE successfully differentiated the positional isomers 2-cyclohexyl-5-cyclopropyl-3-ethylimidazol-4-amine (Fig. 3b) and 5-cyclohexyl-2-cyclopropyl-3-ethylimidazol-4-amine (Fig. 3c) by following distinct inferring paths.

Taken together, our results demonstrate that DiSE achieves high accuracy in structure elucidation across diverse datasets. More importantly, DiSE exhibits interpretability, data efficiency, and expert-like reasoning, offering a practical, scalable, and chemically intuitive framework for automated structure elucidation from spectral data. The model's success stems from both its architectural design and its ability to integrate multiple orthogonal spectral modalities in a coherent and chemically meaningful way.

### 2.4 Structure Elucidation Using Experimental Spectra

To date, DiSE's performance has been demonstrated exclusively on DFT-calculated spectra. To validate its practical utility, we next evaluated DiSE on experimental NMR data, which are inherently more complex due to factors such as temperature, solvent effects, and instrument noise. Because models trained on the QM9-NMR dataset suffered from limited chemical-space coverage and large errors in mPW1PW91-calculated shifts (*52*), we focus here solely on the ensemble DiSE model trained on PCQM4Mv2-NMR. Recognizing that even the more accurate xOPBE functional exhibits non-negligible deviations from experimental chemical shifts, we further applied random perturbations of ±1.0 ppm for $^{13}C$ and ±0.1 ppm for $^{1}H$ chemical shifts (see SM section A.3). Corresponding 2D spectra were reconstructed to reflect these perturbations, thereby improving the model's robustness to real-world application. Detailed results of best model, ensemble models with and without perturbations can be found in SM. Results from three openly available models (CreSS (*39*); CMGNet (*40*) and NMR-resolver (*38*)) were included for comparison.

#### 2.4.1 Structure Elucidation Using Experimental Shifts

Most earlier studies, particularly older ones, report only 1D spectra and lack corresponding 2D spectral data. To address this limitation, we first evaluated DiSE using experimental chemical shifts, with 2D spectra reconstructed from the input 1D chemical shifts and molecular structures. A benchmark set of 45 molecules—36 FDA-approved drugs and 9 intermediates from the Portimine total synthesis (*63*) —was assembled (see Table S5 for Mol ID and Fig. S2 and S3 for detailed structures), spanning simple structure (e.g., Fomepizole) to complex macrocyclic structures (e.g., Portimine A). Notably, 40% of these compounds contain more than 20 heavy atoms, exceeding the size range of the training set and thus challenging DiSE's out-of-distribution generalization (Fig. 4a).

On unperturbed experimental shifts, DiSE correctly identified 38 of 45 structures at Top-1 (84.4% Top-1 accuracy, Fig. 4b), and 42 of 45 within the Top-5 (93.3% Top-5 accuracy; Top-3: 88.9%, Fig. 4c and 4d). All small molecules were placed in the Top-1, while larger or topologically intricate compounds as quantified by the size-normalized spatial score (nSPS, ref. *64*; e.g., dolasetron, nSPS:34.5) appeared within the Top-5 (Fig. 4f and Table S6–11). Molecules bearing exotic motifs—such as the HO–NH–C(=O) group in Vorinostat, which occurs in fewer than 0.01% of training examples —remained challenging but were still recovered by Top-5. Only the most complex molecules (Portimine-18, nSPS:45.7; Portimine A, nSPS: 49.4) were not successfully elucidated, likely due to graph-based resolution limits and insufficiently informative heteroatom features. Introducing random perturbations to chemical shifts modestly improved performance, yielding 86.7% (Top-1), 88.9% (Top-3), and 95.6% (Top-5) accuracy (Table S12). Perturbations also enabled the successful elucidation of molecules previously unresolved. For example, Portimine-7 (Mol ID: 38), which had failed under the ensemble model, was correctly ranked within the Top-4 (Fig. 4f and Table S11). However, for the most complex structures (e.g., Portimine-18 and Portimine A), perturbations alone were insufficient, indicating that additional spectroscopic modalities may be required. We noted that the three openly available models (CReSS, CMGNet and NMR-Solver) were trained exclusively on 1D NMR chemical shifts. Consequently, their performances on this benchmark test were relatively poor, with Top-20 accuracy less than 32% (Table S12). This finding aligns with our ablation studies and previous reports (*12*, *25*), highlighting that 2D NMR spectra are indispensable for automatic as well as accurate structure elucidation.

In summary, these results demonstrate that DiSE—despite being trained solely on calculated chemical shifts—generalizes effectively to experimental chemical shifts. Moreover, the introduction of controlled input noise further enhances predictive performance, underscoring the model's robustness and practical potential. Hence, DiSE can serve as a core component of modern SDLs, enabling autonomous structure elucidation in closed-loop discovery workflows.

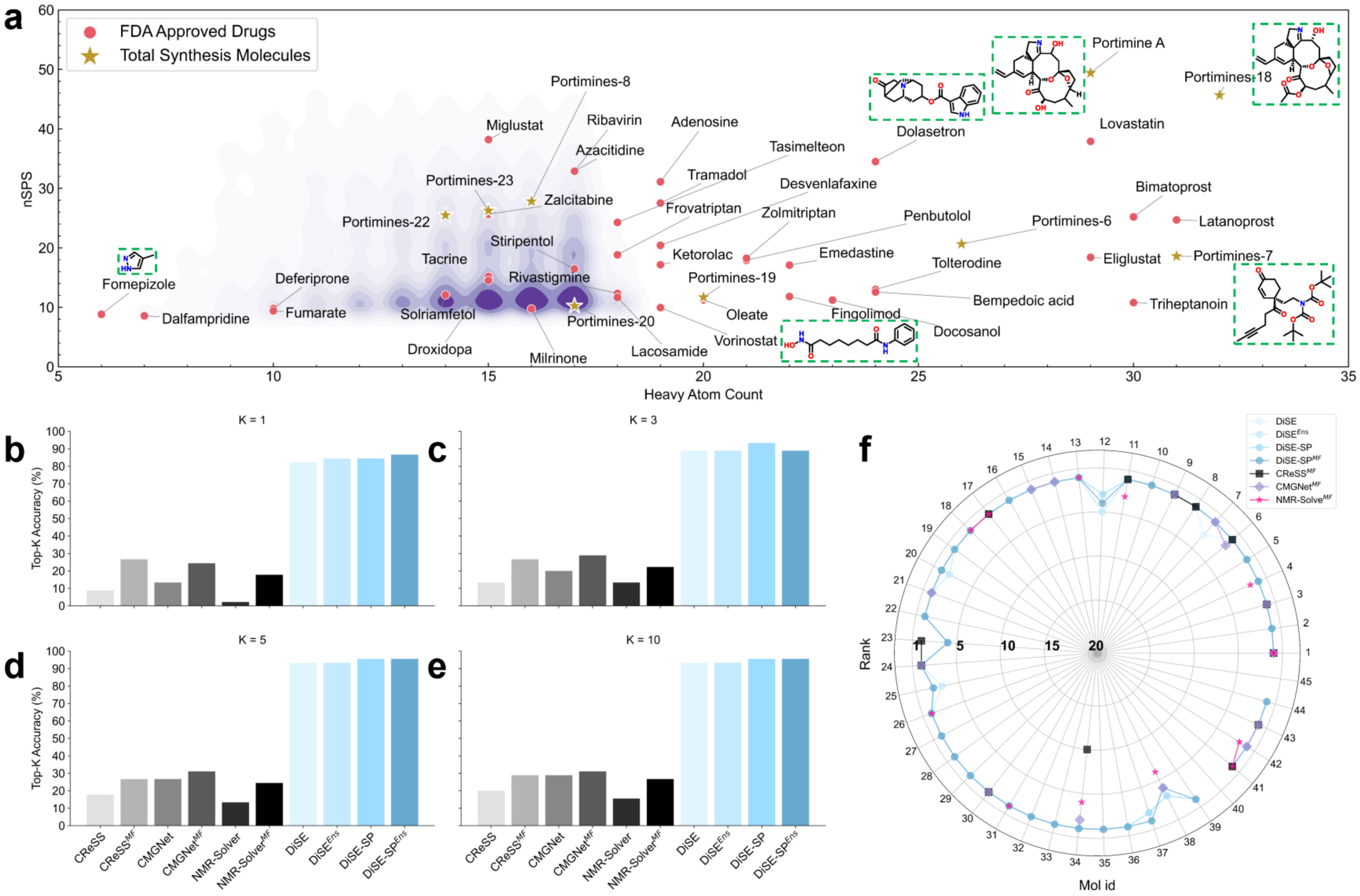

**Fig. 4. Performance of various methods on 45 molecules with experimental chemical shifts. a.** statistic on topological complexity of molecules represented by the size-normalized spatial score (nSPS), where the larger the nSPS value, the more complex the molecule. The statistic on nSPS of training set is represented in purple, where the darker the color, the higher frequency the molecule. **b.- e.** Comparison of CReSS, CMGNet and NMR-Solver with DiSE. Top-1 accuracy (**b**), Top-3 (**c**), Top-5 (**d**) and Top-10 (**e**). **f.** Radar plot illustrating the rank generated for the correct molecular structure by each method for 45 molecules (see Table S5 for Mol ID and Fig. S2 and S3 for structures). Lower ranks (away from the center) indicate better performance. SP refers to small perturbation (±1.0 ppm for $^{13}$C and ±0.1 ppm for $^{1}$H chemical shifts). superscript *Ens* refers to ensemble models and superscript *MF* refers to models with molecular formula filter.

### 2.4.2 Structural Elucidation Using Raw Experimental Spectra

We next assessed DiSE on the most stringent task: *de novo* structure elucidation of natural products from raw experimental data. Eleven recently isolated terpenoid compounds from the soft coral Stereonephthya bellissima (*65*) were selected (see Fig. S4 for detailed structures). To our knowledge, no previous model has attempted *de novo* structure elucidation of natural products of such structurally complexity. These compounds exhibit molecular complexity ranging from 28.52 to 50.7 (Table S13), far exceeding the average molecular complexity 19.12 of the training set, thereby providing a stringent test of DiSE's out-of-distribution generalization and practical applicability.

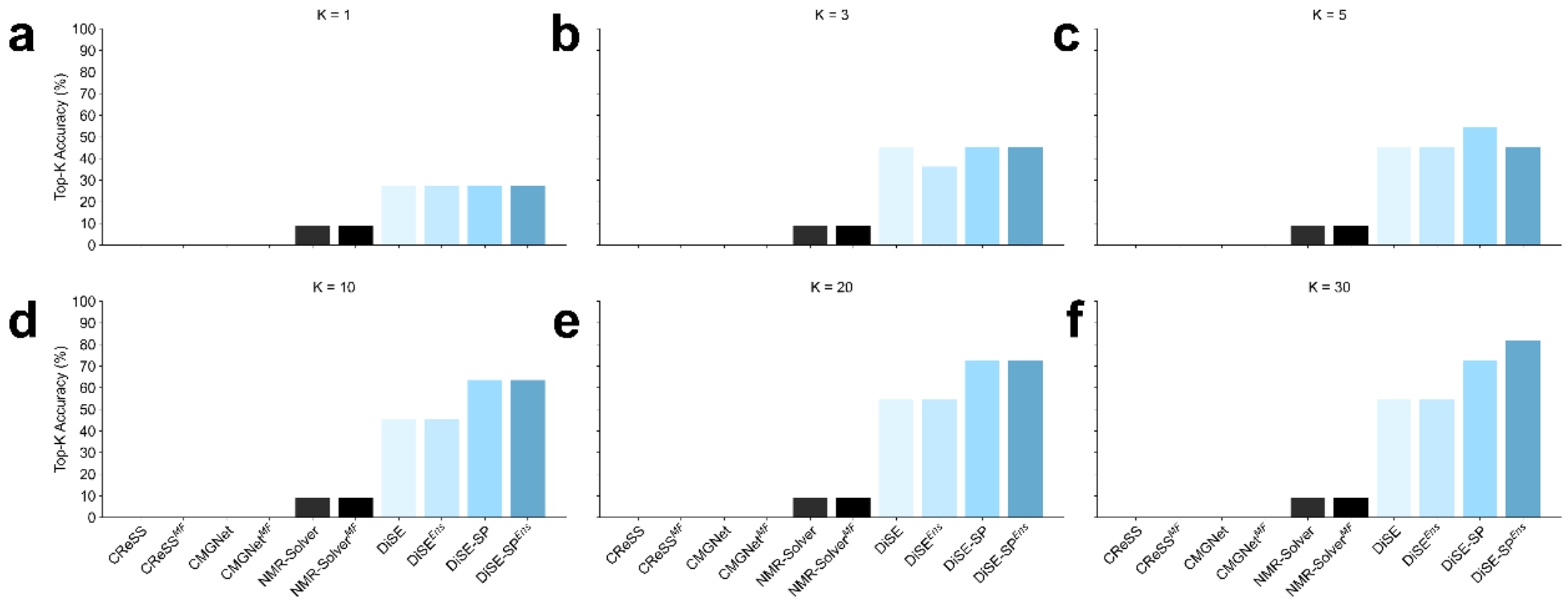

**Fig. 5. *de novo* structure elucidation of natural products from raw experimental data. a.** to **f.** Top-1 accuracy (**a**), Top-3 (**b**), Top-5(**c**), Top-10(**d**), Top-20(**e**) and Top-30(**f**). Detailed structures of the 11 natural products can be found in Fig. S4. SP refers to small perturbation (±1.0 ppm for $^{13}$C and ±0.1 ppm for $^{1}$H chemical shifts). superscript *Ens* refers to ensemble models and superscript *MF* refers to models with molecular formula filter.

As can be seen from Fig. 5, DiSE correctly placed three compounds (bellissinanes 2, 8, and 10) at Top-1 and identified two others (6 and 11) within the Top-5. Expanding to the Top-10 retrieved seven structures (63.6% Top-10 accuracy; 45.5% Top-5), with two additional correct isomers appearing at lower ranks (e.g., bellissinane 1 at rank 22; bellissinane 5 at rank 18), achieving 81.82% Top-30 accuracy. Molecules with highly exotic backbones (bellissinanes 3 and 4) were not recovered. Nevertheless, given the exceptional structural complexity of these natural products, this performance in a blind test is highly encouraging. By contrast, the retrieval-based NMR-Solver resolved only one structure, while CMGNET failed to identify any. With an average runtime of 0.64 seconds per compound on a 4-card RTX 4090 GPU, DiSE demonstrates strong potential for high-throughput, automated structure elucidation in natural product research, drug discovery, and SDLs.

## Conclusion

In this work, we introduced DiSE, an innovative generative diffusion-based architecture for structure elucidation that integrate multiple spectroscopic modality, including MS,

$^{13}C$ and $^{1}H$ chemical shifts, HSQC and COSY. Leveraging graph-based molecular representations, DiSE establishes a probabilistic mapping between spectra and molecular structures. Through this data-driven approach, it achieves high accuracy, generalization, and transferability across chemically diverse datasets. DiSE further demonstrates robustness by maintaining strong performance on experimental data, even when trained solely on calculated spectra. By recovering inherent correlations among multiple spectroscopic modalities, DiSE provides interpretability, data efficiency, and expert-like reasoning, offering a practical, scalable, and chemically intuitive framework for *de novo* structure elucidation. Its capabilities position it as a promising tool with broad applications in natural product discovery, drug development, and self-driving laboratories.

Despite these promising outcomes, several improvements remain. First, DiSE currently handles only neutral molecules composed of C, H, O, and N. Expanding to additional elements would broaden its applicability. Second, Under the current "silent-node" framework, heteroatoms lack dedicated spectral or edge information, limiting accuracy on heteroatom-rich compounds. Incorporating IR and Raman modalities could address this. Furthermore, due to limited edge features ($E_{COSY}$ only), DiSE has difficulties in dealing with molecules with complex topological structures. Integrating additional spectral modalities, such as heteronuclear multiple bond coherence (HMBC) may improve performance in these cases. Finally, DiSE does not yet resolve stereochemistry. Integrating external methods (e.g., SVM-M (*24*), DP4 (*17*)) could enable diastereomer discrimination without fundamental algorithmic changes. By addressing these areas, future versions of DiSE will offer even greater accuracy and broader utility for automated, interpretable structure elucidation. Importantly, there are no fundamental barriers to these developments.

**Acknowledgments**

We gratefully thank both the High-End Computing Center and the CFFF platform of Fudan University for generously providing computational resources. The authors also thank Open Graph Benchmark for providing high-quality equilibrium structures.

**References and Notes**

1. T. Dai *et al.*, *Nature*. **635**, 890–897 (2024).
2. G. Tom *et al.*, *Chem. Rev.* **124**, 9633–9732 (2024).
3. M. Abolhasani, E. Kumacheva, *Nat. Synth*. **2**, 483–492 (2023).
4. M. Seifrid *et al.*, *Acc. Chem. Res.* **55**, 2454–2466 (2022).
5. T. Song *et al.*, *J. Am. Chem. Soc.* **147**, 12534–12545 (2025).
6. C. W. Coley *et al.*, *Science*. **365**, eaax1566 (2019).

7. J. M. Granda, L. Donina, V. Dragone, D.-L. Long, L. Cronin, *Nature*. **559**, 377–381 (2018).

8. V. Sans, L. Porwol, V. Dragone, L. Cronin, *Chem. Sci.* **6**, 1258–1264 (2015).

9. O. Bayley, E. Savino, A. Slattery, T. Noël, *Matter*. **7**, 2382–2398 (2024).

10. J. Liu, J. E. Hein, *Nat. Synth*. **2**, 464–466 (2023).

11. C. J. Taylor et al., Chem. Rev. 123, 3089–3126 (2023).

12. M. Elyashberg, D. Argyropoulos, *Magnetic Reson in Chemistry*. **59**, 669–690 (2021).

13. M. E. Elyashberg, A. Williams, K. Blinov, *The Royal Society of Chemistry*, (2011). https://doi.org/10.1039/9781849734578.

14. D. C. Burns, E. P. Mazzola, W. F. Reynolds, *Nat. Prod. Rep.* **36**, 919–933 (2019).

15. M. Elyashberg, A. Williams, *Molecules*. **26**, 6623 (2021).

16. S. G. Smith, J. M. Goodman, *J. Org. Chem.* **74**, 4597–4607 (2009).

17. S. G. Smith, J. M. Goodman, *J. Am. Chem. Soc.* **132**, 12946–12959 (2010).

18. N. Grimblat, M. M. Zanardi, A. M. Sarotti, *J. Org. Chem.* **80**, 12526–12534 (2015).

19. K. Ermanis, K. E. B. Parkes, T. Agback, J. M. Goodman, *Org. Biomol. Chem.* **15**, 8998–9007 (2017).

20. D. Xin, P.-J. Jones, N. C. Gonnella, *J. Org. Chem.* **83**, 5035–5043 (2018).

21. N. Grimblat, J. A. Gavín, A. Hernández Daranas, A. M. Sarotti, *Org. Lett.* **21**, 4003–4007 (2019).

22. A. Howarth, K. Ermanis, J. M. Goodman, *Chem. Sci.* **11**, 4351–4359 (2020).

23. A. Howarth, J. M. Goodman, *Chem. Sci.* **13**, 3507–3518 (2022).

24. A. Wu *et al.*, *Precision Chemistry*. **1**, 57–68 (2023).

25. M. Priessner *et al.*, ChemRxiv [Preprint] (2025). https://doi.org/10.26434/chemrxiv-2024-zmmnw-v2.

26. M. W. Lodewyk, M. R. Siebert, D. J. Tantillo, *Chem. Rev.* **112**, 1839–1862 (2012).

27. A. V. Buevich, M. E. Elyashberg, *J. Nat. Prod.* **79**, 3105–3116 (2016).

28. Z. Huang, M. S. Chen, C. P. Woroch, T. E. Markland, M. W. Kanan, *Chem. Sci.* **12**, 15329–15338 (2021).

29. F. Hu, M. S. Chen, G. M. Rotskoff, M. W. Kanan, T. E. Markland, *ACS Cent. Sci.* **10**, 2162–2170 (2024).

30. J.-L. Reymond, *Acc. Chem. Res.* **48**, 722–730 (2015).

31. M. A. Yirik, M. Sorokina, C. Steinbeck, *J Cheminform*. **13**, 48 (2021).

32. B. D. McKay, M. A. Yirik, C. Steinbeck, *Journal of Cheminformatics*. **14**, 24 (2022).

33. J. Jumper *et al.*, *Nature*. **596**, 583–589 (2021).

34. Z. Lin *et al.*, *Science*. **379**,1123-1130(2023).

35. J. Abramson *et al.*, *Nature*. **630**, 493–500 (2024).

36. S. Lewis *et al.*, *Science*. **389**, eadv9817 (2025).

37. C. Zeni *et al.*, *Nature*. **639**, 624–632 (2025).

38. Y. Jin *et al.*, arXiv:2509.00640 [physics.chem-ph] (2025). https://doi.org/10.48550/arXiv.2509.00640.

39. Z. Yang *et al.*, *Anal. Chem.* **93**, 16947–16955 (2021).

40. L. Yao *et al.*, *Anal. Chem.* **95**, 5393–5401 (2023).

41. A. Mirza *et al.*, ChemRxiv [Preprint] (2024). https://doi.org/10.26434/chemrxiv-2024-f3b18-v2.

42. M. Alberts, T. Laino, A. C. Vaucher, *Commun Chem*. **7**, 268 (2024).

43. M. Alberts *et al.*, arXiv:2407.17492 [physics.chem-ph] (2024). https://doi.org/10.48550/arXiv.2407.17492.

44. E. Chacko *et al.*, ChemRxiv [Preprint] (2024). https://doi.org/10.26434/chemrxiv-2024-37v2j.

45. T. Hu *et al.*, *J. Am. Chem. Soc.* **147**, 27525–27536 (2025).

46. L. Wang *et al.*, arXiv:2507.06853 [cs.LG] (2025). https://doi.org/10.48550/arXiv.2507.06853.

47. I. Shumailov *et al.*, *Nature*. **631**, 755–759 (2024).

48. C. Vignac *et al.*, arXiv:2209.14734 [cs.LG] (2023). https://doi.org/10.48550/arXiv.2209.14734.

49. H. Chen, T. Liang, K. Tan, A. Wu, X. Lu, *J Cheminform*. **16**, 132 (2024).

50. A. Gupta, S. Chakraborty, R. Ramakrishnan, *Mach. Learn.: Sci. Technol.* **2**, 035010 (2021).

51. M. Nakata, T. Shimazaki, *J. Chem. Inf. Model.* **57**, 1300–1308 (2017).

52. J. Zhang, Q. Ye, C. Yin, A. Wu, X. Xu, *J. Phys. Chem. A.* **124**, 5824–5831 (2020).

53. RDKit: Open-source cheminformatics. https://www.rdkit.org.

54. M. Bohde *et al.*, arXiv:2502.09571 [cs.LG] (2025). https://doi.org/10.48550/arXiv.2502.09571.

55. Z. Tu, C. W. Coley, *J. Chem. Inf. Model.* **62**, 3503–3513 (2022).

56. J. F. Joung *et al.*, *Angew Chem Int Ed*, e202411296 (2024).

57. P. Schwaller, T. Gaudin, D. Lányi, C. Bekas, T. Laino, *Chem. Sci.* **9**, 6091–6098 (2018).

58. M. Priessner *et al.*, *J. Chem. Inf. Model.* **64**, 3180–3191 (2024).

59. J. Sohl-Dickstein *et al.*, arXiv:1503.03585 [cs.LG] (2015). https://doi.org/10.48550/arXiv.1503.03585.

60. Y. Song *et al.*, arXiv:2011.13456 [cs.LG] (2021). https://doi.org/10.48550/arXiv.2011.13456.

61. J. Ho *et al.*, arXiv:2006.11239 [cs.LG] (2020). https://doi.org/10.48550/arXiv.2006.11239.

62. D. Bajusz, A. Rácz, K. Héberger, *J Cheminform*. **7**, 20 (2015).

63. J. Tang *et al.*, *Nature*. **622**, 507–513 (2023).

64. A. Krzyzanowski, A. Pahl, M. Grigalunas, H. Waldmann, *J. Med. Chem.* **66**, 12739–12750 (2023).

65. X. Yu et al., J. Nat. Prod. 87, 1150–1158 (2024).

# Supplementary Materials

## A Materials and Methods

### A.1. Formal Molecular Graph Representation

In the DiSE framework, each molecule, including noise molecules, is encoded as an undirected graph *G*, defined as:

$$G = (X, E, E_{COSY}, Y) \qquad \text{(S1)}$$

Where *X* represents node features, *E* represents edge features, $E_{COSY}$ represents COSY-specific edges, and *Y* represents global features.

#### *A.1.1. Node features X*

The node feature tensor, *X*, is a concatenation of the following components:

$$X = \left[ X_{type}, X_{shifts}, X_{cycles}, X_{\lambda}, X_{valence}, X_{charge} \right] \qquad \text{(S2)}$$

$X_{type}$ (Atom Type): A one-hot encoded vector identifying the atom or super-atom type. The representation varies depending on the available spectroscopic data:

1. For input data derived from molecular formula and $^{13}$C and/or $^{1}$H NMR, the atoms are encoded as C, H, O, N (4 types).

2. For input data derived from molecular formula and HSQC, the atoms are encoded as super-atoms —such as $CH_x$ (where $x \in \{0, 1, 2, 3\}$), $OH_y$ (where $y \in \{0, 1\}$), and $NH_z$ (where $z \in \{0, 1, 2\}$), resulting in 9 distinct node types.

$X_{Shifts}$ (Chemical Shifts): Encodes the chemical shift of the carbon atom and those of any attached protons. For methylene ($CH_2$) groups, the average shifts of the two hydrogen atoms are used. A padding value of 0 is applied for atoms lacking attached protons (e.g., $CH_0$) and exchangeable protons (OH/NH/$NH_2$).

$X_{cycles}$ (Ring Membership, *48*, *66*): A three-element vector encoding the atom's membership in 3-, 4-, and 5-membered rings.

$X_{\lambda}$ (Eigenvalues Features, *48*, *66*): This feature consists of the following components:

1. A binary indicator for whether the node is part of the largest connected component of the graph.
2. The eigenvectors corresponding to the two smallest non-zero eigenvalues of the graph Laplacian matrix.

$X_{valence}$ (Valence State): Represents the valence state of an atom, calculated by summing the bond orders of all incident edges. Bond orders are defined as 1.0, 1.5, 2.0, and 3.0 for single, aromatic, double, and triple bonds, respectively.

$X_{charge}$ (Formal Charge): Represents the formal charge of each atom. It is calculated as the difference between the expected valence of the atom or super-atom (e.g., 4 for $CH_0$, 3 for CH, 2 for $CH_2$) and its calculated valence state, $X_{valence}$.

For $X_{spectra}$, it is the concatenation of $X_{types}$ and $X_{shifts}$.

*A.1.2. Edge Features E*

In the molecular graph, edges represent the covalent bonds between pairs of atoms. The feature vector $e_{ij}$ for an edge connecting atoms or super-atoms *i* and *j* is a K-dimensional one-hot tensor that encodes the bond type. The supported bond types are {No-Bond, Single, Double, Triple, Aromatic, Single-Aromatic}.

*A.1.3. COSY Edge Features* $E_{COSY}$

To fully leverage the other 2D NMR, we explicitly incorporate information from $^3J_{HH}$ COSY spectra. A COSY correlation provides strong evidence of a three-bond connectivity between two vicinal protons (H−$C_i$−$C_j$−H). In our graph representation, this corresponds to a known bond between the respective carbon atoms (nodes $C_i$ and $C_j$).

This information is integrated as a distinct channel in the edge feature representation. We define a binary matrix $\boldsymbol{M}_{\mathrm{COSY}}$ where an element $(M_{\mathrm{COSY}})_{ij} = 1$ if a $^3J_{HH}$ correlation is observed between the protons on nodes *i* and *j*, and $(M_{\mathrm{COSY}})_{ij} = 0$ otherwise. This matrix is then concatenated with the standard edge feature tensor $E_{zero}$, which is padded with zeros to match the dimensions of $\boldsymbol{M}_{\mathrm{COSY}}$, creating an augmented edge representation.

This allows the model to distinguish between edges that are known with high certainty from the COSY data and those that must be inferred. This known subgraph serves as a strong constraint during the denoising phase of the diffusion model.

*A.1.4.* Global Features *Y*

The global feature tensor *Y* is composed of four components that are concatenated:

1. $y_{cycles}$ (Cycle Counts, *48*, *66*): The total count of cycles with lengths from 3 to 6 within the graph.
2. $n_{components}$ (Connected Components, *48, 66*): The number of connected components in the graph.
3. Laplacian Eigenvalues (*48, 66*): The five smallest non-zero eigenvalues of the graph's Laplacian matrix.
4. Diffusion Timestep: The normalized diffusion timestep, *t*.

A.2. Noise-injection Stage and Denoising Stage

We define the noise model as *q* and the denoising model (neural network) as $\varphi_\theta$ .

*A.2.1.* Noise-injection Stage

The training process involves a forward Markov chain that progressively corrupts an initial edge tensor, $E_0$, producing the sequence ($E_1$, $E_2$, ..., $E_{Tmax}$).
The level of noise is determined by a predefined noise schedule. This schedule uses a cumulative alpha value $\bar{\alpha}_t$ based on the cosine function to define:

$$\bar{\alpha}_t = \frac{f(t)}{f(0)} \qquad \text{(S3)}$$

$$f(t)=\cos(\frac{\pi}{2}\cdot\frac{(t/T_{\max})+s}{1+s})^2 \quad \text{(S4)}$$

Here, $s$ is a small offset, default set is 0.008. From $\bar{\alpha}_t$, we define the single-step noise level $\beta_t$ as:

$$\beta_t = 1-\alpha_t = 1-\frac{\bar{\alpha}_t}{\bar{\alpha}_{t-1}} \quad \text{(S5)}$$

The transition from state $E_{t-1}$ to $E_t$ is controlled by a one-step transition matrix $Q_t$, which applies noise:

$$q(E_t \mid E_{t-1}) = E_{t-1}\cdot Q_t \quad \text{(S6)}$$

$$Q_t = (1-\beta_t)\cdot I + \beta_t\cdot K = \alpha_t\cdot I + (1-\alpha_t)\cdot K \quad \text{(S7)}$$

Where $I$ is the identity matrix, $K$ is the distribution matrix of training set edge types (see section A.1.2.).

According to the Markov property, the probability of transitioning from the initial state $E_0$ to any state $E_t$ is determined by the cumulative effect of the transition matrix at each step. This cumulative transition is represented by the matrix $\bar{Q}_t$ :

$$\bar{Q}_t = \prod_{i=1}^{t} Q_i = \bar{\alpha}_t\cdot I + (1-\bar{\alpha}_t)\cdot K \quad \text{(S8)}$$

#### *A.2.2.* Denoising Stage

When training $\varphi_\theta$, optimize the cross-entropy loss function $L$ between the predicted probability $E_p$ and the true edge $E_0$:

$$L = \sum_{1\le i,j\le n} cross-entropy(e_{i,j}^{True}, e_{i,j}^{Pred}) \quad \text{(S9)}$$

When performing inference, first construct $E_{Tmax}$: perform multinomial sampling on the distribution of edges in the training set to obtain a random noise graph. Let the trained network perform denoising on this noise graph. It will predict a slightly clearer, more structurally robust graph $G_{Tmax-1}$. Use this newly generated, slightly cleaner graph $G_{Tmax-1}$ as input again, and have the network continue to denoise it by one step, resulting in $G_{Tmax-2}$.

Repeat this process, with each step making the graph clearer and more structured, until we finally obtain a completely clean and structurally sound new graph $G_0$ (this is pred structure).

### A.3. Model Evaluation Strategies

#### *A.3.1.* Model Inference and Evaluation Strategies

We define two different types of metrics from model inference: best and ensemble models.

Best Model: This result corresponds to a single prediction based on the best model identified from the N inference runs.

Ensemble Model: The result is the sum of the best and second-best models.

In the DiSE test set evaluation, we made N = 100 independent inferences for each molecule. In the detailed case studies, the number of independent inferences per molecule was increased to N = 256 to ensure more comprehensive structural sampling (N=128 for perturbation inference).

### *A.3.2.* Model Robustness and Perturbation Inference

To assess the stability of the model and robustness to noise in the experimental data, we performed a perturbation inference. We introduced random noise to the input chemical shifts by adding a value sampled from a uniform distribution, [-δ, +δ], to the experimental shifts.

Small Perturbation (SP): $\delta_C$ = 1.0 ppm ($^{13}C$ shifts) and $\delta_H$ = 0.1 ppm ($^{1}H$ shifts)

Medium Perturbation (MP): $\delta_C$ = 3.0 ppm ($^{13}C$ shifts) and $\delta_H$ = 0.5 ppm ($^{1}H$ shifts)

Large Perturbation (LP): $\delta_C$ = 5.0 ppm ($^{13}C$ shifts) and $\delta_H$ = 1.0 ppm ($^{1}H$ shifts)

**Table S1. Summary of deep generative models for structure elucidation.**

| Acronym | Architecture | Molecular Representation | Data Source | Input Spectra |
|---|---|---|---|---|
| SGG (28) | CNN | Functional group | MestReNova[a] | MS + 1D NMR |
| MT (29) | CNN + Transformer | Functional group | MestReNova[a] | 1D NMR |
| MMST (25) | Transformer | SMILES | SGNN[b] | MS + 1D/2D NMR + IR |
| CMGNet (40) | BART | SMILES | DFT + Exp | MS + $^{13}$C NMR |
| S2S (41) | Transformer | SMILES | MestReNova[a] | MS + 1D NMR |
| MST (43) | Transformer | SMILES | MestReNova[a] | MS + 1D NMR |
| TranSpec (45) | CNN + Transformer | SMILES | DFT | IR / Raman |

[a] Data were generated using the MestreNova software package (*67*).

[b] Data were generated using the SGNN (Scalable graph neural network) model (*68*).

Abbreviations:
SGG, Substructure Graph Generator;
MT, Multitask Transformer;
MMST, MultiModalSpectralTransformer;
CMGNet, Conditional Molecular Generation Net;
S2S, Spec2Struct;
MST, Multimodal Spectroscopic Transformer;
CNN, Convolutional Neural Network;
1D/2D NMR, One-/Two-Dimensional Nuclear Magnetic Resonance;
BART, Bidirectional and Auto-Regressive Transformers;
SMILES, Simplified Molecular-Input Line-Entry System;
DFT, Density Functional Theory;
Exp, Experimental data;
IR, Infrared spectroscopy.

**Table S2. Hyperparameter configurations for the DiSE models.**

| Hyperparameter | DiSE-QM9-NMR | DiSE-PCQM4Mv2-NMR | DiSE-QM9-NMR-ablation |
|---|---|---|---|
| diffusion steps | 500 | 500 | 500 |
| diffusion noise schedule | custom | custom | cosine |
| n_layers | 24 | 20 | 12 |
| hidden_mlp_dims_X | 256 | 1024 | 256 |
| hidden_mlp_dims_E | 128 | 512 | 128 |
| hidden_mlp_dims_y | 128 | 512 | 128 |
| hidden_dims_dx | 256 | 1024 | 256 |
| hidden_dims_de | 64 | 256 | 64 |
| hidden_dims_dy | 64 | 256 | 64 |
| hidden_dims_n_head | 8 | 32 | 8 |
| hidden_dims_dim_ffX | 256 | 1024 | 256 |
| hidden_dims_dim_ffE | 128 | 512 | 128 |
| hidden_dims_dim_ffy | 128 | 512 | 128 |
| edge_types_no_bond | 7.26e-1 | 8.50e-1 | 7.26e-1 |
| edge_types_single_bond | 2.24e-1 | 9.72e-2 | 2.24e-1 |
| edge_types_double_bond | 1.85e-2 | 8.63e-3 | 1.85e-2 |
| edge_types_triple_bond | 8.70e-3 | 8.98e-4 | 8.70e-3 |
| edge_types_aromatic_bond | 2.29e-2 | 4.28e-2 | 2.29e-2 |
| edge_types_single_aromatic_bond | \ | 8.28e-4 | \ |
| learning rate | 2.00e-3 | 2.00e-3 | 2.00e-3 |
| weight_decay | 1.00e-12 | 1.00e-12 | 1.00e-12 |

**Table S3. Top-K accuracy (%) of DiSE variants on the QM9-NMR and PCQM4Mv2-NMR test sets.**

| | Top-K accuracy (%) | | | |
|---|---|---|---|---|
| methods | 1 | 3 | 5 | 10 |
| DiSE-QM9-NMR-Best | 92.76 | 96.78 | 97.17 | 97.26 |
| DiSE-QM9-NMR-Ensemble | 93.59 | 98.22 | 98.62 | 98.79 |
| DiSE-PCQM4Mv2-NMR-Best | 92.11 | 96.41 | 97.03 | 97.44 |
| DiSE-PCQM4Mv2-NMR-Ensemble | 92.55 | 96.95 | 97.42 | 98.14 |

**Table S4. Top-K accuracy (%) of DiSE using different combinations of spectroscopic inputs.**

| methods | Top-K accuracy (%) | | | | |
|---|---|---|---|---|---|
| | 1 | 3 | 5 | 10 | All |
| MS + $^{1}$H NMR | 13.93 | 27.78 | 35.16 | 44.10 | 63.55 |
| MS + $^{13}$C NMR | 20.19 | 34.77 | 40.68 | 47.56 | 61.03 |
| MS + $^{1}$H & $^{13}$C NMR | 34.93 | 50.76 | 55.85 | 61.63 | 68.19 |
| MS + $^{1}$H & $^{13}$C NMR + HSQC | 80.68 | 90.34 | 91.70 | 92.41 | 92.65 |
| MS + $^{1}$H & $^{13}$C NMR + HSQC + COSY | 90.66 | 96.26 | 96.76 | 96.94 | 96.99 |
| MS + $^{1}$H* & $^{13}$C NMR + HSQC + COSY | 92.22 | 96.54 | 96.84 | 96.94 | 96.94 |

superscript *: exchangeable protons information used

**Table S5. Names, molecular formulas, nSPS and references for 45 molecules (36 FDA-approved drugs and 9 intermediates from the Portimine total synthesis).**

| No. | Name | MF | nSPS | Ref. |
|---|---|---|---|---|
| 1 | Ribavirin | $C_8H_{12}N_4O_5$ | 32.88 | (*69*) |
| 2 | Lovastatin | $C_{24}H_{36}O_5$ | 37.9 | (*70*, *71*) |
| 3 | Penbutolol | $C_{18}H_{29}NO_2$ | 17.9 | (*72*) |
| 4 | Milrinone | $C_{12}H_9N_3O$ | 9.75 | (*73*) |
| 5 | Oleate | $C_{18}H_{34}O_2$ | 11.25 | (*74*) |
| 6 | Adenosine | $C_{10}H_{13}N_5O_4$ | 31.11 | (*75*) |
| 7 | Ketorolac | $C_{15}H_{13}NO_3$ | 17.16 | (*76*) |
| 8 | Zalcitabine | $C_9H_{13}N_3O_3$ | 25.67 | (*77*, *78*) |
| 9 | Tacrine | $C_{13}H_{14}N_2$ | 15.2 | (*79*) |
| 10 | Tramadol | $C_{16}H_{25}NO_2$ | 27.53 | (*80*) |
| 11 | Latanoprost | $C_{26}H_{40}O_5$ | 24.71 | (*81*, *82*) |
| 12 | Dolasetron | $C_{19}H_{20}N_2O_3$ | 34.5 | (*83*) |
| 13 | Zolmitriptan | $C_{16}H_{21}N_3O_2$ | 18.24 | (*84*) |
| 14 | Fomepizole | $C_4H_6N_2$ | 8.83 | (*77*, *85*) |
| 15 | Emedastine | $C_{17}H_{26}N_4O$ | 17.09 | (*86*, *87*) |
| 16 | Tolterodine | $C_{22}H_{31}NO$ | 13 | (*88*) |
| 17 | Rivastigmine | $C_{14}H_{22}N_2O_2$ | 12.33 | (*89*) |
| 18 | Docosanol | $C_{22}H_{46}O$ | 11.22 | (*90*) |
| 19 | Bimatoprost | $C_{25}H_{37}NO_4$ | 25.2 | (*82*, *91*) |
| 20 | Frovatriptan | $C_{14}H_{17}N_3O$ | 18.83 | (*92*) |
| 21 | Miglustat | $C_{10}H_{21}NO_4$ | 38.2 | (*93*) |
| 22 | Azacitidine | $C_8H_{12}N_4O_5$ | 32.88 | (*90*) |
| 23 | Vorinostat | $C_{14}H_{20}N_2O_3$ | 9.95 | (*94*) |
| 24 | Desvenlafaxine | $C_{16}H_{25}NO_2$ | 20.42 | (*95*) |
| 25 | Lacosamide | $C_{13}H_{18}N_2O_3$ | 11.67 | (*96*) |
| 26 | Dalfampridine | $C_5H_6N_2$ | 8.57 | (*90*) |
| 27 | Fingolimod | $C_{19}H_{33}NO_2$ | 11.82 | (*97*) |
| 28 | Deferiprone | $C_7H_9NO_2$ | 9.8 | (*90*) |
| 29 | Fumarate | $C_6H_8O_4$ | 9.4 | (*90*) |
| 30 | Tasimelteon | $C_{15}H_{19}NO_2$ | 24.28 | (*98*) |
| 31 | Droxidopa | $C_9H_{11}NO_5$ | 14.53 | (*99*, *100*) |
| 32 | Eliglustat | $C_{23}H_{36}N_2O_4$ | 18.41 | (*101*) |
| 33 | Stiripentol | $C_{14}H_{18}O_3$ | 16.47 | (*102*) |
| 34 | Solriamfetol | $C_{10}H_{14}N_2O_2$ | 12.07 | (*103*, *104*) |
| 35 | Bempedoic acid | $C_{19}H_{36}O_5$ | 12.58 | (*105*) |
| 36 | Triheptanoin | $C_{24}H_{44}O_6$ | 10.8 | (*106*) |
| 37 | Portimines-6 | $C_{19}H_{31}NO_6$ | 20.65 | (*63*) |
| 38 | Portimines-7 | $C_{24}H_{35}NO_6$ | 18.61 | (*63*) |
| 39 | Portimines-8 | $C_{14}H_{17}NO$ | 27.81 | (*63*) |
| 40 | Portimines-18 | $C_{25}H_{33}NO_6$ | 45.66 | (*63*) |
| 41 | Portimines-19 | $C_{14}H_{25}NO_5$ | 11.7 | (*63*) |
| 42 | Portimines-20 | $C_{13}H_{15}NO_3$ | 10.24 | (*63*) |
| 43 | Portimines-22 | $C_{10}H_{16}O_4$ | 25.5 | (*63*) |
| 44 | Portimines-23 | $C_{11}H_{20}O_4$ | 26.27 | (*63*) |
| 45 | Portimine A | $C_{23}H_{31}NO_5$ | 49.41 | (*63*) |

**Table S6.** Ranking performance of DiSE for the 36 FDA-approved drugs using chemical shifts from minimum-energy conformation ($\boldsymbol{\delta_{ME}}$), conformationally averaged shifts ($\boldsymbol{\delta_{CA}}$), and experimental data ($\boldsymbol{\delta_{exp}}$). Results are shown for the single best model and the ensemble model. The numeral indicates the rank of the correct structure, and the symbol "F" indicates that the correct structure was not found.

| | | $\delta_{ME}$-rank | | $\delta_{CA}$-rank | | $\delta_{exp}$-rank | |
|---|---|---|---|---|---|---|---|
| No. | Name | Best | Ensemble | Best | Ensemble | Best | Ensemble |
| 1 | Ribavirin | 1 | 1 | 1 | 1 | 1 | 1 |
| 2 | Lovastatin | 1 | 1 | 2 | 2 | 1 | 1 |
| 3 | Penbutolol | 1 | 1 | 1 | 1 | 1 | 1 |
| 4 | Milrinone | 1 | 1 | 1 | 1 | 1 | 1 |
| 5 | Oleate | 1 | 1 | 1 | 1 | 1 | 1 |
| 6 | Adenosine | 1 | 1 | 1 | 1 | 1 | 1 |
| 7 | Ketorolac | 1 | 1 | 1 | 1 | 3 | 1 |
| 8 | Zalcitabine | 1 | 1 | 1 | 1 | 1 | 1 |
| 9 | Tacrine | 1 | 1 | 1 | 1 | 1 | 1 |
| 10 | Tramadol | 1 | 1 | 1 | 1 | 1 | 1 |
| 11 | Latanoprost | 1 | 1 | 1 | 1 | 1 | 1 |
| 12 | Dolasetron | F | 1 | F | F | 5 | 5 |
| 13 | Zolmitriptan | 1 | 1 | 1 | 1 | 1 | 1 |
| 14 | Fomepizole | 1 | 1 | 1 | 1 | 1 | 1 |
| 15 | Emedastine | 1 | 1 | 1 | 1 | 1 | 1 |
| 16 | Tolterodine | 1 | 1 | 1 | 1 | 1 | 1 |
| 17 | Rivastigmine | 1 | 1 | 1 | 1 | 1 | 1 |
| 18 | Docosanol | 1 | 1 | 1 | 1 | 1 | 1 |
| 19 | Bimatoprost | 1 | 1 | 1 | 1 | 1 | 1 |
| 20 | Frovatriptan | 1 | 1 | 1 | 1 | 1 | 2 |
| 21 | Miglustat | 1 | 1 | 1 | 1 | 1 | 1 |
| 22 | Azacitidine | 2 | 2 | 2 | 2 | 1 | 1 |
| 23 | Vorinostat | 2 | 1 | 2 | 2 | 4 | 4 |
| 24 | Desvenlafaxine | 1 | 1 | 1 | 1 | 1 | 1 |
| 25 | Lacosamide | 1 | 1 | 1 | 1 | 2 | 3 |
| 26 | Dalfampridine | 1 | 1 | 1 | 1 | 1 | 1 |
| 27 | Fingolimod | 1 | 1 | 1 | 1 | 1 | 1 |
| 28 | Deferiprone | 1 | 1 | 1 | 1 | 1 | 1 |
| 29 | Fumarate | 1 | 1 | 1 | 1 | 1 | 1 |
| 30 | Tasimelteon | 1 | 1 | 1 | 1 | 1 | 1 |
| 31 | Droxidopa | 1 | 1 | 1 | 1 | 1 | 1 |
| 32 | Eliglustat | 1 | 1 | 1 | 1 | 1 | 1 |
| 33 | Stiripentol | 1 | 1 | 1 | 1 | 1 | 1 |
| 34 | Solriamfetol | 1 | 1 | 1 | 1 | 1 | 1 |
| 35 | Bempedoic acid | 1 | 1 | 1 | 1 | 1 | 1 |
| 36 | Triheptanoin | 1 | 1 | 1 | 1 | 1 | 1 |

**Table S7.** Ranking performance of DiSE for the 9 intermediates from the Portimine total synthesis using experimental shifts ($\boldsymbol{\delta_{exp}}$). Results are shown for the single best model and the ensemble model. The numeral indicates the rank of the correct structure, and the symbol "F" indicates that the correct structure was not found.

| | | $\delta_{ME}$-rank | | $\delta_{CA}$-rank | | $\delta_{exp}$-rank | |
|---|---|---|---|---|---|---|---|
| No. | Name | Best | Ensemble | Best | Ensemble | Best | Ensemble |
| 1 | Portimines-6 | - | - | - | - | 2 | 1 |
| 2 | Portimines-7 | - | - | - | - | F | F |
| 3 | Portimines-8 | - | - | - | - | 1 | 1 |
| 4 | Portimines-18 | - | - | - | - | F | F |
| 5 | Portimines-19 | - | - | - | - | 1 | 1 |
| 6 | Portimines-20 | - | - | - | - | 1 | 1 |
| 7 | Portimines-22 | - | - | - | - | 1 | 1 |
| 8 | Portimines-23 | - | - | - | - | 1 | 1 |
| 9 | Portimine A | - | - | - | - | F | F |

**Table S8.** Ranking performance of CReSS, CMGNet, and NMR-Solver for the 36 FDA-approved drugs using experimental shifts ($\boldsymbol{\delta_{exp}}$). Results are shown for runs with (MF) and without (NoMF) molecular formula information. The numeral indicates the rank of the correct structure, and the symbol "F" indicates that the correct structure was not found.

| No. | Name | CReSS-rank | | CMGNet-rank | | NMR-Solver-rank | |
|---|---|---|---|---|---|---|---|
| | | NoMF | MF | NoMF | MF | NoMF | MF |
| 1 | Ribavirin | F | F | 1 | 1 | F | F |
| 2 | Lovastatin | F | F | F | F | 14 | 1 |
| 3 | Penbutolol | F | F | F | F | F | F |
| 4 | Milrinone | F | F | F | F | F | F |
| 5 | Oleate | 12 | 10 | 2 | 2 | 13 | 4 |
| 6 | Adenosine | 4 | 1 | 3 | 1 | F | F |
| 7 | Ketorolac | 1 | 1 | F | F | F | F |
| 8 | Zalcitabine | 1 | 1 | 7 | 1 | F | F |
| 9 | Tacrine | F | F | 4 | 1 | F | F |
| 10 | Tramadol | F | F | F | F | F | 1 |
| 11 | Latanoprost | F | F | F | F | F | F |
| 12 | Dolasetron | F | F | F | F | 10 | 2 |
| 13 | Zolmitriptan | F | F | F | F | F | F |
| 14 | Fomepizole | 5 | 1 | 1 | 1 | 3 | 1 |
| 15 | Emedastine | F | F | F | F | F | F |
| 16 | Tolterodine | F | F | F | F | F | F |
| 17 | Rivastigmine | F | F | 4 | 1 | 3 | 2 |
| 18 | Docosanol | 8 | 1 | F | F | 2 | 1 |
| 19 | Bimatoprost | F | F | F | F | F | F |
| 20 | Frovatriptan | F | F | F | F | F | F |
| 21 | Miglustat | 20 | 1 | 1 | 1 | F | F |
| 22 | Azacitidine | 1 | 1 | F | 2 | F | F |
| 23 | Vorinostat | F | F | 2 | 1 | F | F |
| 24 | Desvenlafaxine | F | F | F | F | F | F |
| 25 | Lacosamide | 1 | 1 | F | F | F | F |
| 26 | Dalfampridine | 2 | 1 | 1 | 1 | F | F |
| 27 | Fingolimod | F | F | F | F | F | F |
| 28 | Deferiprone | 15 | 1 | F | F | 1 | 1 |
| 29 | Fumarate | F | 1 | 1 | 1 | F | F |
| 30 | Tasimelteon | F | F | F | F | 3 | 1 |
| 31 | Droxidopa | F | F | 1 | 1 | F | F |
| 32 | Eliglustat | F | F | F | F | F | F |
| 33 | Stiripentol | F | F | F | F | 16 | 1 |
| 34 | Solriamfetol | F | F | F | F | F | F |
| 35 | Bempedoic acid | F | F | F | F | F | F |
| 36 | Triheptanoin | F | F | F | F | F | F |

**Table S9.** Ranking performance of CReSS, CMGNet, and NMR-Solver for the 9 intermediates from the Portimine total synthesis using experimental shifts ($\boldsymbol{\delta_{exp}}$). Results are shown for runs with (MF) and without (NoMF) molecular formula information. The numeral indicates the rank of the correct structure, and the symbol "F" indicates that the correct structure was not found.

| | | CReSS-rank | | CMGNet-rank | | NMR-Solver-rank | |
|---|---|---|---|---|---|---|---|
| No. | Name | NoMF | MF | NoMF | MF | NoMF | MF |
| 1 | Portimines-6 | F | F | F | F | F | F |
| 2 | Portimines-7 | F | F | F | F | F | F |
| 3 | Portimines-8 | F | F | F | F | F | F |
| 4 | Portimines-18 | F | F | F | F | F | F |
| 5 | Portimines-19 | F | F | F | F | F | F |
| 6 | Portimines-20 | F | F | F | F | F | F |
| 7 | Portimines-22 | 1 | 1 | F | F | 2 | 1 |
| 8 | Portimines-23 | F | F | 5 | 4 | F | 6 |
| 9 | Portimine A | F | F | F | F | F | F |

**Table S10.** Ranking performance of DiSE for the 36 FDA-approved drugs using experimental chemical shifts with varying levels of perturbations: small (DiSE-SP), medium (DiSE-MP), and large (DiSE-LP). Results are shown for the single best model and the ensemble model. The numeral indicates the rank of the correct structure.

| | | DiSE-SP-rank | | DiSE-MP-rank | | DiSE-LP-rank | |
|---|---|---|---|---|---|---|---|
| No. | Name | Best | Ensemble | Best | Ensemble | Best | Ensemble |
| 1 | Ribavirin | 1 | 1 | 1 | 1 | 1 | 1 |
| 2 | Lovastatin | 1 | 1 | 2 | 2 | 6 | 4 |
| 3 | Penbutolol | 1 | 1 | 1 | 1 | 1 | 1 |
| 4 | Milrinone | 1 | 1 | 1 | 1 | 1 | 1 |
| 5 | Oleate | 1 | 1 | 1 | 1 | 1 | 1 |
| 6 | Adenosine | 1 | 1 | 1 | 1 | 1 | 1 |
| 7 | Ketorolac | 1 | 1 | 1 | 1 | 1 | 1 |
| 8 | Zalcitabine | 1 | 1 | 1 | 1 | 1 | 1 |
| 9 | Tacrine | 1 | 1 | 1 | 1 | 1 | 1 |
| 10 | Tramadol | 1 | 1 | 1 | 1 | 1 | 1 |
| 11 | Latanoprost | 1 | 1 | 1 | 1 | 1 | 1 |
| 12 | Dolasetron | 3 | 4 | 1 | 1 | 1 | 1 |
| 13 | Zolmitriptan | 1 | 1 | 1 | 1 | 6 | 1 |
| 14 | Fomepizole | 1 | 1 | 1 | 1 | 1 | 1 |
| 15 | Emedastine | 1 | 1 | 1 | 1 | 1 | 1 |
| 16 | Tolterodine | 1 | 1 | 1 | 1 | 2 | 2 |
| 17 | Rivastigmine | 1 | 1 | 1 | 1 | 1 | 1 |
| 18 | Docosanol | 1 | 1 | 1 | 1 | 1 | 1 |
| 19 | Bimatoprost | 1 | 1 | 1 | 1 | 1 | 1 |
| 20 | Frovatriptan | 1 | 1 | 1 | 1 | 1 | 2 |
| 21 | Miglustat | 1 | 1 | 1 | 1 | 1 | 1 |
| 22 | Azacitidine | 1 | 1 | 1 | 1 | 1 | 1 |
| 23 | Vorinostat | 4 | 4 | 4 | 2 | 4 | 2 |
| 24 | Desvenlafaxine | 1 | 1 | 1 | 1 | 1 | 1 |
| 25 | Lacosamide | 2 | 2 | 2 | 3 | 1 | 4 |
| 26 | Dalfampridine | 1 | 1 | 1 | 1 | 1 | 1 |
| 27 | Fingolimod | 1 | 1 | 1 | 1 | 1 | 1 |
| 28 | Deferiprone | 1 | 1 | 1 | 1 | 1 | 1 |
| 29 | Fumarate | 1 | 1 | 1 | 1 | 1 | 1 |
| 30 | Tasimelteon | 1 | 1 | 1 | 1 | 1 | 1 |
| 31 | Droxidopa | 1 | 1 | 1 | 1 | 1 | 1 |
| 32 | Eliglustat | 1 | 1 | 1 | 1 | 1 | 1 |
| 33 | Stiripentol | 1 | 1 | 1 | 1 | 1 | 1 |
| 34 | Solriamfetol | 1 | 1 | 1 | 1 | 1 | 1 |
| 35 | Bempedoic acid | 1 | 1 | 1 | 1 | 1 | 1 |
| 36 | Triheptanoin | 1 | 1 | 1 | 1 | 1 | 1 |

**Table S11.** Ranking performance of DiSE for the 9 intermediates from the Portimine total synthesis using experimental chemical shifts with varying levels of perturbations: small (DiSE-SP), medium (DiSE-MP), and large (DiSE-LP). Results are shown for the single best model and the ensemble model. The numeral indicates the rank of the correct structure, and the symbol "F" indicates that the correct structure was not found.

| | | DiSE-SP-rank | | DiSE-MP-rank | | DiSE-LP-rank | |
|---|---|---|---|---|---|---|---|
| No. | Name | Best | Ensemble | Best | Ensemble | Best | Ensemble |
| 1 | Portimines-6 | 2 | 1 | 1 | 1 | 8 | 1 |
| 2 | Portimines-7 | 3 | 4 | 2 | 2 | F | F |
| 3 | Portimines-8 | 1 | 1 | 1 | 1 | 1 | 1 |
| 4 | Portimines-18 | F | F | F | F | F | F |
| 5 | Portimines-19 | 1 | 1 | 1 | 1 | 1 | 1 |
| 6 | Portimines-20 | 1 | 1 | 1 | 1 | 1 | 1 |
| 7 | Portimines-22 | 1 | 1 | 1 | 1 | 1 | 1 |
| 8 | Portimines-23 | 1 | 1 | 1 | 1 | 1 | 1 |
| 9 | Portimine A | F | F | F | F | F | F |

**Table S12.** Comparative Top-K accuracy (%) of various methods on the 45 molecules benchmark set. Performance is shown for alternative methods (CReSS, CMGNet, NMR-Solver) with (MF) and without (NoMF) molecular formula, and for DiSE under different conditions: using unperturbed experimental shifts (DiSE-exp), and experimental shifts with small (DiSE-SP), medium (DiSE-MP), and large (DiSE-LP) perturbations. Results are reported for the best model (B) and ensemble model (E).

| | Top-K accuracy (%) | | | | |
|---|---|---|---|---|---|
| methods | 1 | 3 | 5 | 10 | 20 |
| CReSS-NoMF | 11.11 | 13.33 | 17.78 | 20.00 | 26.67 |
| CReSS-MF | 26.67 | 26.67 | 26.67 | 28.89 | 28.89 |
| CMGNet-NoMF | 13.33 | 20.00 | 26.67 | 28.89 | 28.89 |
| CMGNet-MF | 24.44 | 28.89 | 31.11 | 31.11 | 31.11 |
| NMR-Solver-NoMF | 2.22 | 13.33 | 13.33 | 15.56 | 22.22 |
| NMR-Solver-MF | 17.78 | 22.22 | 24.44 | 26.67 | 26.67 |
| DiSE-exp-B | 82.22 | 88.89 | 93.33 | 93.33 | 93.33 |
| DiSE-exp-E | 84.44 | 88.89 | 93.33 | 93.33 | 93.33 |
| DiSE-SP-B | 84.44 | 93.33 | 95.56 | 95.56 | 95.56 |
| DiSE-SP-E | **86.67** | 88.89 | **95.56** | **95.56** | **95.56** |
| DiSE-MP-B | 86.67 | 93.33 | 95.56 | 95.56 | 95.56 |
| DiSE-MP-E | 86.67 | **95.56** | 95.56 | 95.56 | 95.56 |
| DiSE-LP-B | 82.22 | 84.44 | 86.67 | 93.33 | 93.33 |
| DiSE-LP-E | 82.22 | 88.89 | 93.33 | 93.33 | 93.33 |

**Table S13.** Names, molecular formulas, nSPS and references for 11 isolated terpenoid compounds from the soft coral Stereonephthya Bellissima.

| No. | Name | MF | nSPS | Ref. |
|---|---|---|---|---|
| 1 | Bellissinanes-1 | $C_{17}H_{26}O_3$ | 50.70 | (*65*) |
| 2 | Bellissinanes-2 | $C_{17}H_{28}O_3$ | 38.65 | (*65*) |
| 3 | Bellissinanes-3 | $C_{23}H_{29}NO_2$ | 33.85 | (*65*) |
| 4 | Bellissinanes-4 | $C_{23}H_{27}NO$ | 28.52 | (*65*) |
| 5 | Bellissinanes-5 | $C_{14}H_{20}O_3$ | 52.24 | (*65*) |
| 6 | Bellissinanes-6 | $C_{14}H_{22}O_3$ | 46.06 | (*65*) |
| 7 | Bellissinanes-7 | $C_{14}H_{24}O_4$ | 52.78 | (*65*) |
| 8 | Bellissinanes-8 | $C_{20}H_{34}O_2$ | 39.18 | (*65*) |
| 9 | Bellissinanes-9 | $C_{22}H_{36}O_3$ | 36.04 | (*65*) |
| 10 | Bellissinanes-10 | $C_{20}H_{36}O_2$ | 38.09 | (*65*) |
| 11 | Bellissinanes-11 | $C_{21}H_{36}O_3$ | 35.58 | (*65*) |

**Fig. S1. The general architecture of the Graph Transformer block.**

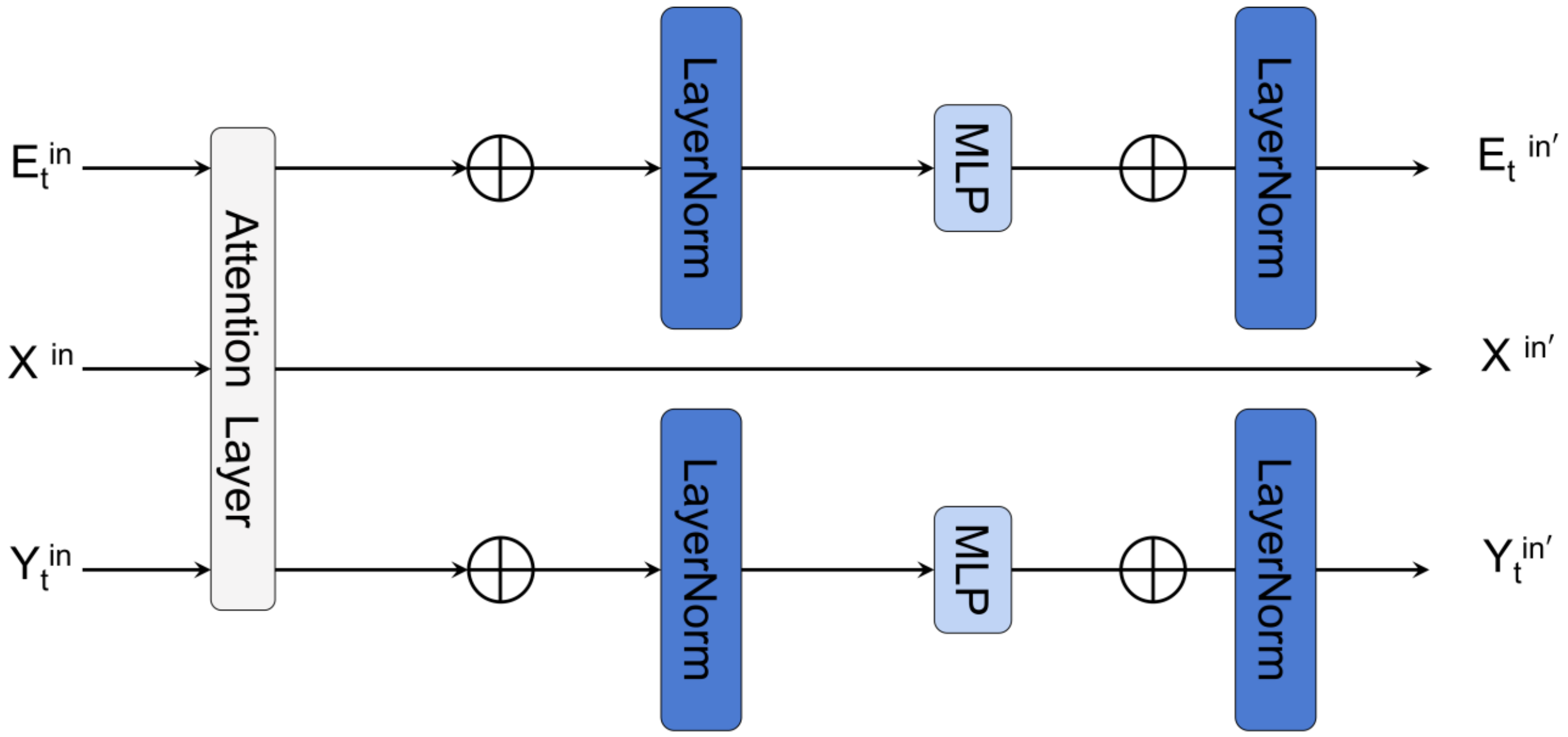

# Fig. S2. Molecular Structures of 36 FDA-approved drugs.

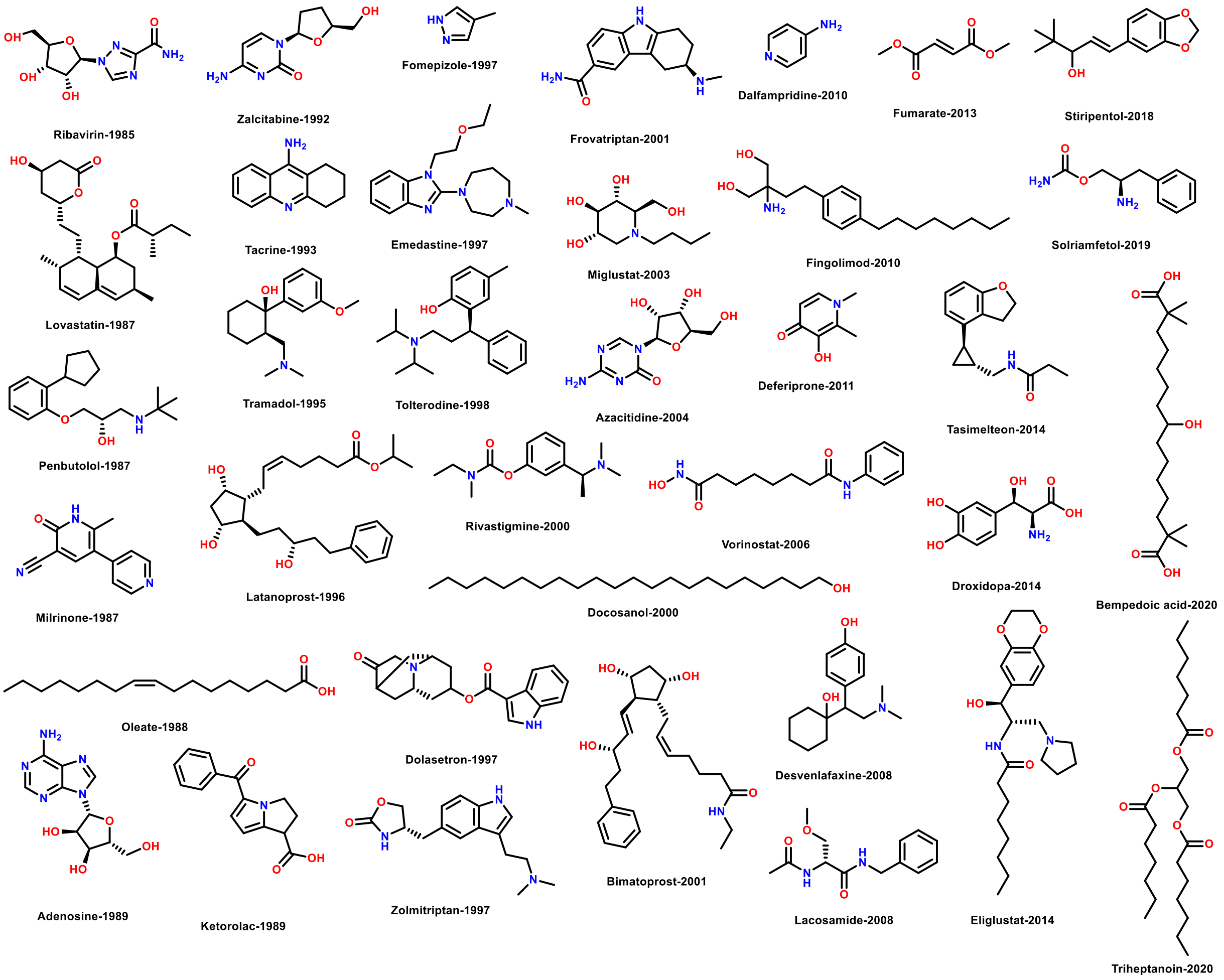

**Fig. S3. Molecular Structures of 9 intermediates from the Portimine total synthesis.**

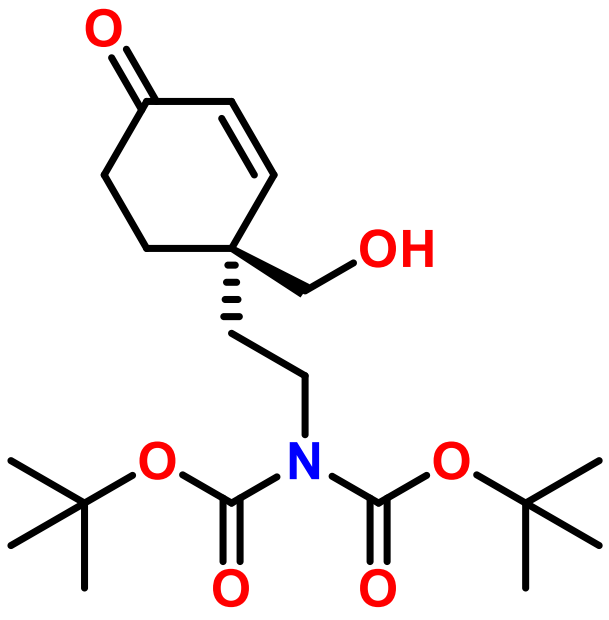

Portimines-6

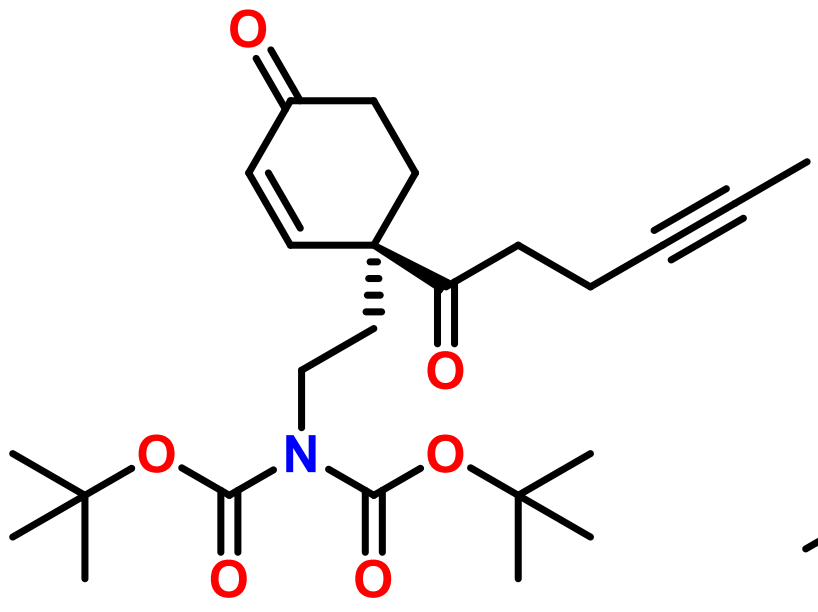

Portimines-7

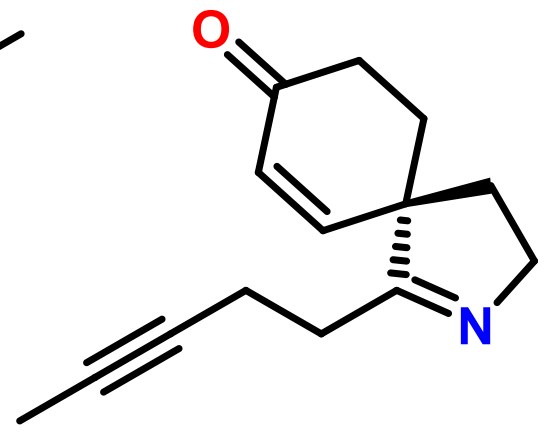

Portimines-8

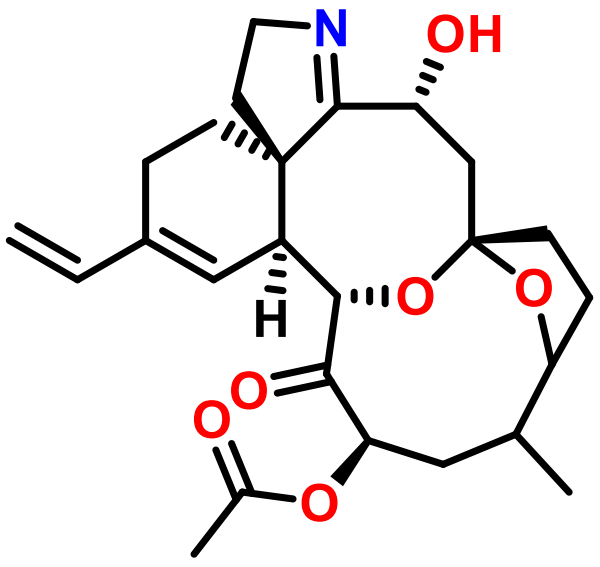

Portimines-18

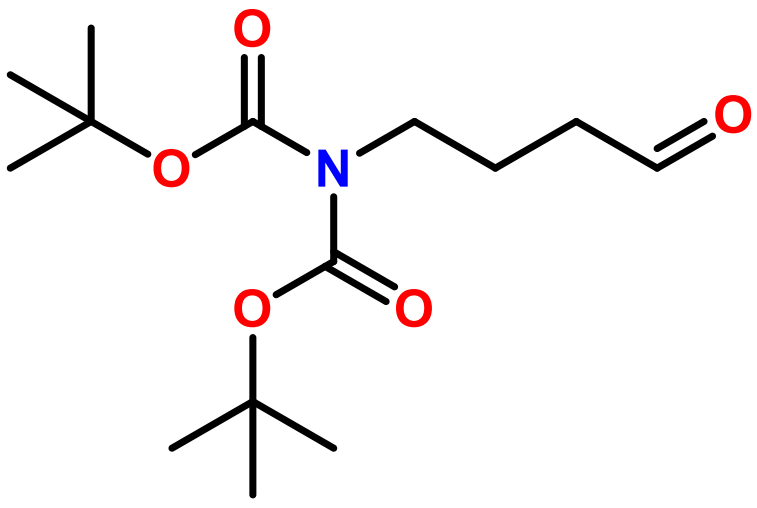

Portimines-19

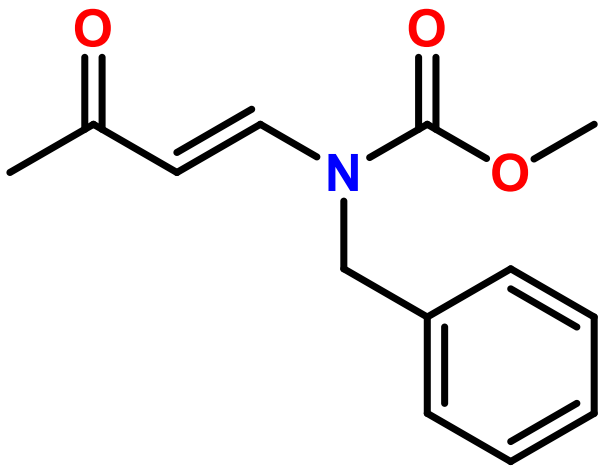

Portimines-20

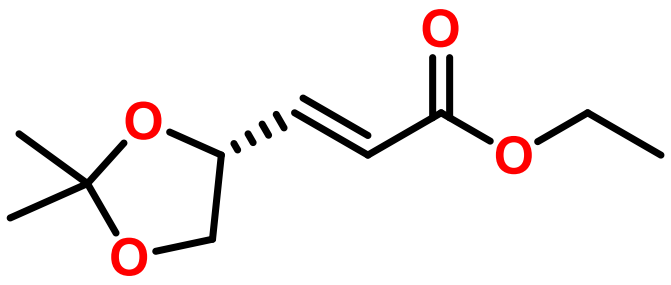

Portimines-22

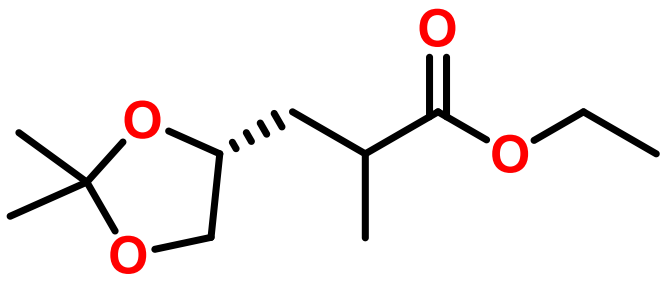

Portimines-23

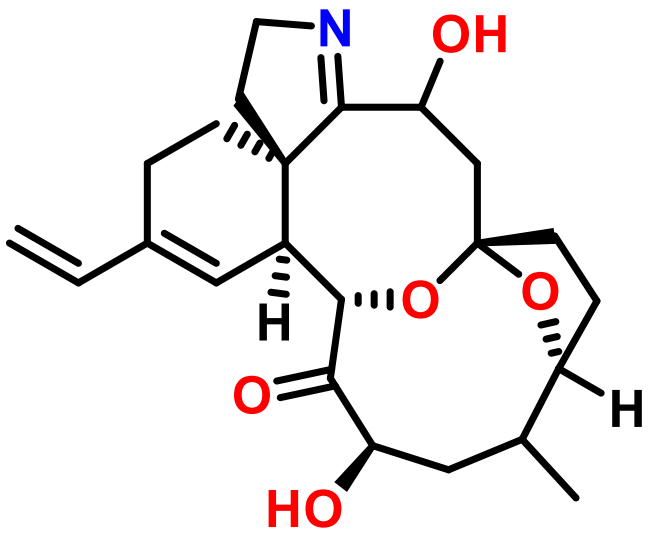

Portimine-A

**Fig. S4.** Molecular Structures of 11 isolated terpenoid compounds from the soft coral Stereonephthya Bellissima.

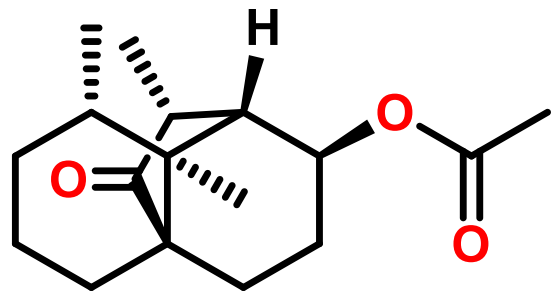

**Bellissinanes-1**

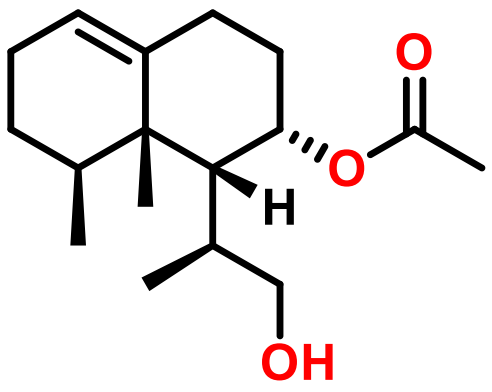

**Bellissinanes-2**

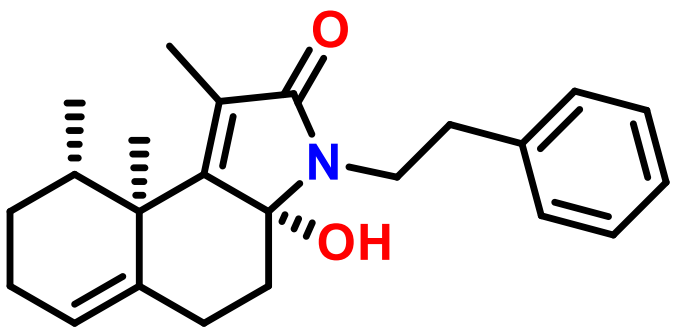

**Bellissinanes-3**

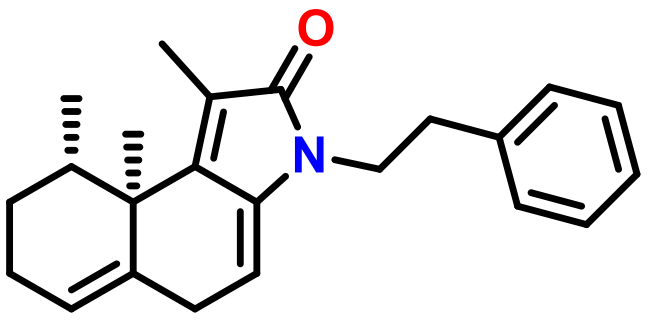

**Bellissinanes-4**

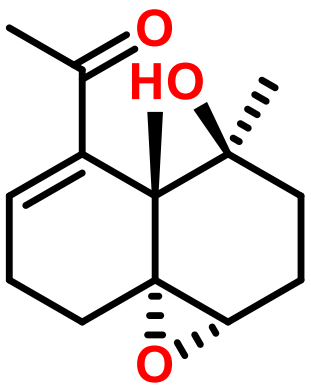

**Bellissinanes-5**

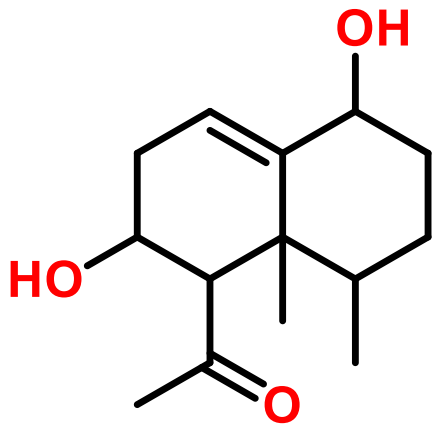

**Bellissinanes-6**

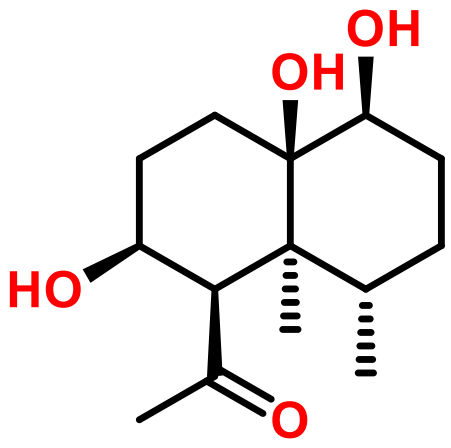

**Bellissinanes-7**

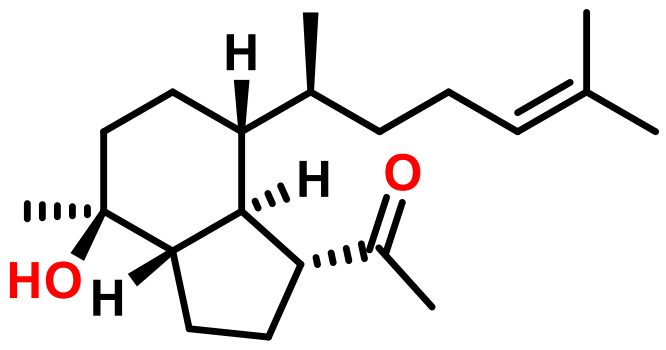

**Bellissinanes-8**

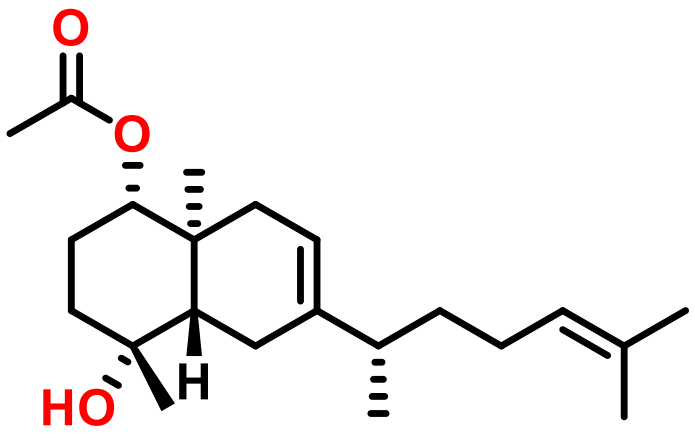

**Bellissinanes-9**

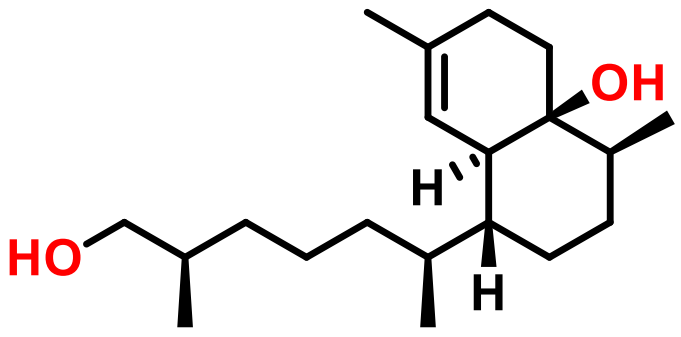

**Bellissinanes-10**

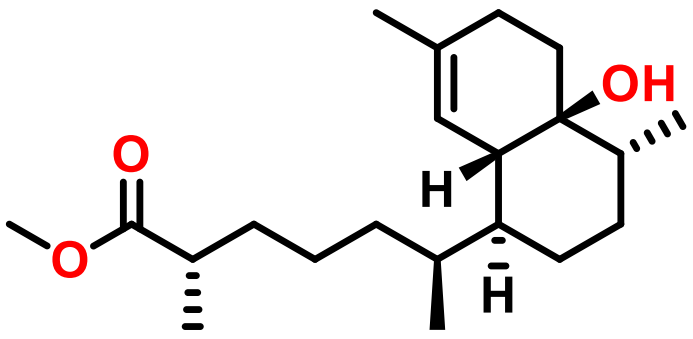

**Bellissinanes-11**